\documentclass[12pt]{iopart}

\usepackage{graphicx}
\usepackage{caption}
\usepackage{subcaption}
\usepackage{placeins}
\usepackage{float}
\usepackage{lineno}
\usepackage{hyperref}

\begin{document}

\title[]{Determining neutral fueling response to RMPs in MAST using a multi-reservoir particle balance model and EMC3‑EIRENE}

\author{Kurt Flesch$^{1,2}$, James Harrison$^3$, Andrew Kirk$^3$, Ian Waters$^4$, Heinke Frerichs$^1$, Oliver Schmitz$^1$, Livia Casali$^2$}

\address{$^1$University of Wisconsin-Madison, 1500 Engineering Dr., Madison, WI 53706}
\address{$^2$University of Tennessee-Knoxville, 823 Neyland Dr., Knoxville, TN 37916}
\address{$^3$CCFE Fusion Association, Culham Science Centre, Abingdon, Oxfordshire, OX14 3DB, UK}
\address{$^4$Lawrence Livermore National Laboratory}
\ead{kflesch@utk.edu}
\vspace{10pt}
\begin{indented}
\item[]
\end{indented}

\begin{abstract}
The application of certain configurations of resonant magnetic perturbations (RMPs) has been shown to cause a plasma pump-out in both L- and H-mode discharges at MAST \cite{kirk_resonant_2010}. In this paper we discuss the impact of neutral fueling on this density reduction. The neutral fueling and average particle confinement time $\tau_p$ of the main ion species were calculated using a 0-D particle balance analysis. When the RMPs were applied, it was found that there was an increase in ionizations and a 15$\%$ reduction in $\tau_p$ for L-mode discharges and a similar result for inter-ELM (edge localized modes) periods of H-mode discharges. A time-dependent global multi-reservoir particle balance (MRPB) was developed, which included atomic and molecular reservoirs, to further investigate the role neutrals had on the density change. We discuss how this model was able to accurately reproduce the experimentally measured density reduction and ionization increase due to either a reduction in $\tau_p$ or a reduction in particle fueling efficiency. Results from EMC3-EIRENE modeling indicate this change could be attributed to neutral particle fueling occurring in locations with now-opened field lines due to the chaotic edge-region from RMP applications. 
\end{abstract}

%
%
%
%
%

\section{\label{sec:intro}Introduction}
For successful operation, ITER \cite{loarte_new_2025} and future fusion power plants will need to achieve simultaneously high core performance with an acceptable boundary solution characterized by low divertor temperature, $T_{target}<5$ eV, low steady-state heat load on the divertor target, $q_\perp<10$ MW/m$^2$, and very limited levels of intermittent heat flux \cite{loarte_effects_2001}\cite{kallenbach_impurity_2013}. The integration of the hot core with the cold edge, also known as the "core-edge integration," remains one of the biggest challenges in magnetic confinement fusion \cite{casali_improved_2020}\cite{turco_physics_2023}\cite{kallenbach_partial_2015}\cite{casali_achievement_2025}. The standard tokamak operation mode is the H-mode regime \cite{wagner_quarter-century_2007}, which features high performance in the core due to the build-up of an edge transport barrier. However, in H-mode operation, transient MHD events called Edge Localized Modes (ELMs) \cite{zohm_edge_1996} result in large heat and particle losses, and extrapolating from current experiments, type I ELMs will melt or destroy the plasma facing components and cannot be allowed to occur \cite{loarte_transient_2007} \cite{martin_design_2013}. This requires either mitigating these ELMs to a manageable level or completely suppressing them. One method of ELM mitigation and suppression is to use external coils to produce resonant magnetic perturbations (RMPs). This has been shown to be effective at numerous experiments like DIII-D \cite{evans_suppression_2004}, JET \cite{liang_active_2007}, ASDEX-U \cite{suttrop_first_2011}, KSTAR \cite{jeon_suppression_2012}, and EAST \cite{sun_first_2021}. At MAST, the RMPs were found to mitigate the ELMs but not fully suppress them \cite{kirk_resonant_2010}. In H-mode, when the RMPs are set in a configuration that results in ELM mitigation at MAST, there is also a reduction in density, the \textit{plasma pump-out} \cite{kirk_resonant_2010}. This plasma pump-out can also occur in L-mode configurations, and has been observed in all of the experiments with ELM suppression and mitigation.

In order to explain the cause of this RMP-induced pump-out, extensive work has been done to study the effect these RMPs have on transport, both in the core and edge. At some devices, changes in the radial electric field have been observed, which may change the $E \times B$ shear profile \cite{mordijck_radial_2014}. Changes in the turbulence have also been observed in both experiment \cite{mckee_increase_2013}\cite{wilcox_evidence_2016} and simulations \cite{wilcox_modeling_2017} that could impact the particle transport. Non-linear MHD modeling has shown that neoclassical ion diffusion may also play a role in the increased transport \cite{kim_transition_2023}. Recent results at KSTAR show that kink-like modes appear when RMPs are applied that cause non-symmetric plasma surface displacements, which can also enhance neoclassical transport \cite{lee_observation_2025}. At MAST, there was evidence of an increase in turbulence near the edge in L-mode scenarios \cite{tamain_edge_2010} but not in H-mode. Altogether, there is evidence that there are changes in transport during RMP application, but there is currently no cohesive mechanism across the many devices that have conducted these RMPs experiments \cite{jakubowski_uence_2014}. Transport, however, is only one side of the balance that determines plasma density. Plasma particles are refueled by neutral atoms and molecules that are ionized at the plasma edge, either from recycled neutrals or gas puffing, or from particles added to the core using NBI or pellet fueling. Therefore, fueling from neutrals plays an important role in establishing plasma density alongside the particle transport out of the plasma. It is well known that the application of the RMPs changes the magnetic field structure in the edge, and the edge region is where these fueling neutrals become ionized. If the magnetic structure changes, then the location and efficiency of the particle fueling may also change. The work in this paper will address changes to the effectiveness of the neutral fueling at the edge as the RMPs are applied.

This paper will be comprised of 5 sections discussing the methods and results. The first section outlines the RMP experiments conducted at MAST that will be used for analysis. This includes a description of both L-mode and H-mode discharges that exhibit the density pump-out. Section \ref{sec:partconfine} introduces a global particle balance as a first order approximation to quantify the impact of RMPs on both the fueling and the exhaust to the plasma, treated as a single reservoir. Section \ref{sec:multires} extends the particle balance to multiple reservoirs, which now includes a reservoir for each of the atomic and molecular neutral populations along with the plasma. In Section \ref{sec:EMC3}, modeling using the edge plasma and neutral code EMC3-EIRENE is presented along with an argument for a new 3-D treatment for the scrape-off layer (SOL). Conclusions from this work and next steps to take are discussed in Section \ref{sec:con}.
 
\section{\label{sec:experiment}Experimental Setup}

\begin{figure}
\centering
    \begin{subfigure}{0.45\textwidth}
        \includegraphics[width=\textwidth]{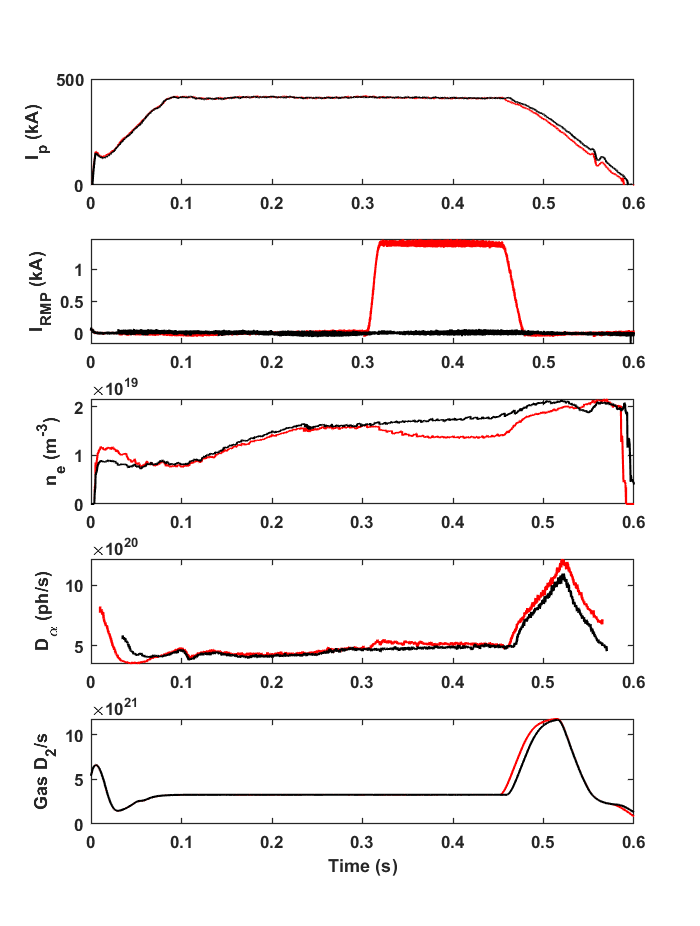}
        \caption{L-mode}
        \label{fig:L-mode_parameters}
    \end{subfigure}
    \begin{subfigure}{0.45\textwidth}
        \includegraphics[width=\textwidth]{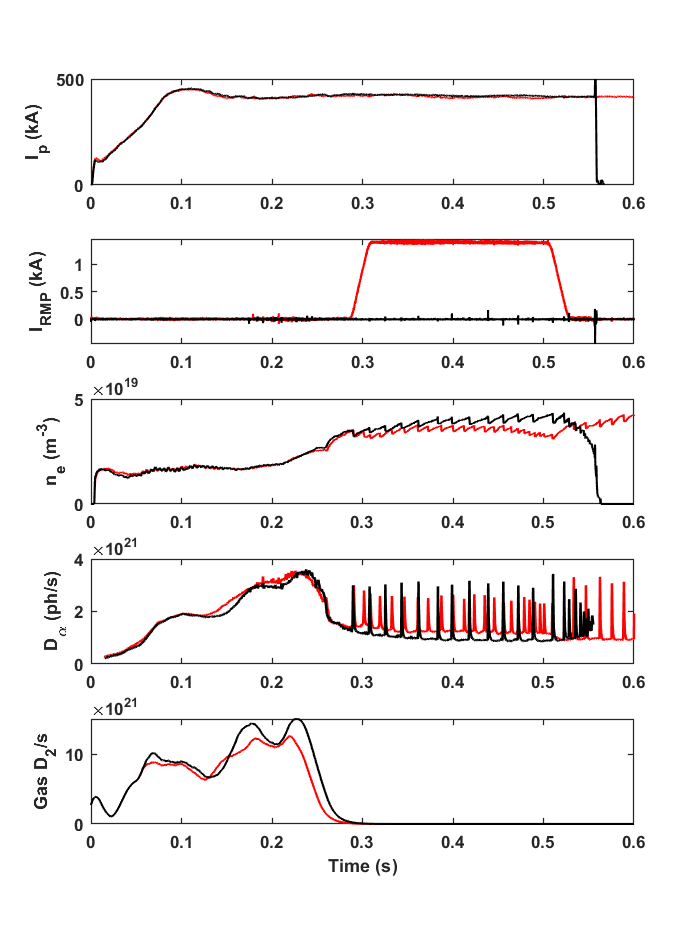}
        \caption{H-mode}
        \label{fig:H-mode_parameters}
    \end{subfigure} 
    \caption{Time evolution of two L-mode (21711 and 21712) and two H-mode (29092 and 29178) discharges, the axisymmetric discharges are shown in black and RMP discharges are red.}
    \label{fig:exp_parameters}
\end{figure}

The Mega-Ampere Sphecical Tokamak (MAST) was a spherical tokamak with a major radius of $0.9$m and minor radius of $\sim0.6$m. It had a maximum plasma current of $1.3$MA and maximum toroidal field of $0.55$T. In this paper, results from two L-mode and two H-mode scenarios will be discussed. The first scenarios were connected double null (CDN) L-mode discharges, shots 21711 and 21712. These discharges had Ohmic heating with $I_p=400$kA, $B_T=0.52$T, and $q_{95}=6.2$, as well as constant neutral gas puffing on the outer edge during the plasma flat-top. The $\beta_N$ started at 0.65 for both and went up to 0.8 for the axisymmetric case and down to 0.6 for the RMP. Time traces of the plasma current, RMP coil current, average electron density, $D_\alpha$ emission, and gas puffing can be seen in Figure \ref{fig:L-mode_parameters}. For the L-mode scenario, n=3 RMPs are applied at ~0.3s, as indicated by the coil current, and the subsequent pump-out begins nearly simultaneously, seen in the plasma density. There is also an accompanying spike in $D_\alpha$ at that time.

The other experimental scenario used for this paper is a lower single null (LSN) H-mode, shots 29092 and 29178, that had neutral beam heating with n=6 RMPs applied, $I_p=400kA$, $B_T=0.52$T, and $q_{95}$ changing from 5.5 to 3.5 during the time of interest. The $\beta_N$ started at 4.0 for both and then went up to 4.5 for the axisymmetric case and down to 3.5 for the RMP. A similar set of time profiles for the H-mode discharges are shown in Figure \ref{fig:H-mode_parameters}.  In this, the neutral gas puffing is shut off as the plasma transitions to H-mode, and the RMP coils are turned on just after 0.3s. As the RMPs are turned on, the ELMs are mitigated, seen with the increased frequency of $D_\alpha$ spikes and smaller density drops. Also just as the RMP coils reach their peak current, the average inter-ELM density can be seen to decrease, indicating pump-out, and the inter-ELM $D_\alpha$ emission increases.

\section{\label{sec:partconfine}Single reservoir particle balance}

As mentioned in the first section, the plasma density is established by both the sources and sinks, where the source terms come from neutral ionizations fueling the plasma and the sink terms come from transport processes removing particles from the plasma. By simplifying the plasma as a single reservoir, the following ansatz can be used to begin to quantify those terms.
\begin{eqnarray}
    \frac{dN}{dt} = -\frac{N}{\tau_p} + \Phi_i
    \label{eqn:one}
\end{eqnarray}
where $N$ is the total number of particles of the main ion species in the reservoir, $\tau_p$ is the average confinement time for the main ion species, and $\Phi_i$ is the fueling rate from ionized neutrals. The term $N/\tau_p$ is the rate particles leave the plasma, and in this, the many complicated transport terms are simplified down to a single value in the average particle confinement time. To first order, this allows one to determine the total impact of the changing terms of transport and fueling as the RMPs are applied.

Assuming ambipolarity and deuterium as the main ion species, the total number of ion particles $N$ can be calculated by multiplying the line-averaged density as measured by the interferometer and the plasma volume as calculated from the equilibrium reconstruction solver EFIT. The source term $\Phi_i$ is estimated from measured quantities using the following method:
\begin{eqnarray}
    \Phi_i = 4\pi \cdot A_p \cdot \frac{S}{XB} I_{D_\alpha}
    \label{eqn:two}
\end{eqnarray}
where $A_p$ is the surface area of the outboard side of the plasma from the top of the plasma to the X-point along the separatrix as calculated from EFIT, $S/XB$ is the branching ratio of the number of ionization events that occur per number of $D_\alpha$ photons collected, and $I_{D_\alpha}$ is the $D_\alpha$ emission ($photons\cdot sr^{-1}m^{-2}s^{-1}$). The $D_\alpha$ emission used in the calculations is measured by an absolutely calibrated filterscope that views tangentially through the edge of the plasma at the midplane.  The view of the filterscope excludes light from the divertor regions so that the fueling calculation only includes the ionizations in plasma edge near the last closed flux surface because ions born here are more likely to fuel the plasma than those in the divertor region.

\begin{figure}
\centering
    \begin{subfigure}{0.45\textwidth}
        \includegraphics[width=\textwidth]{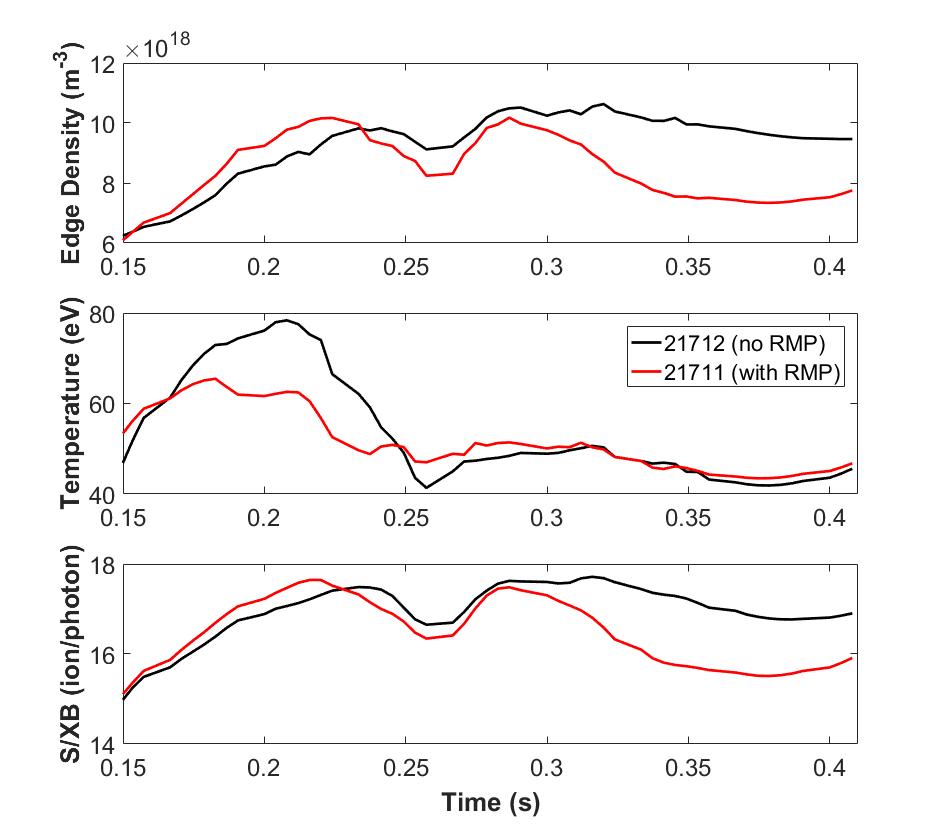}
        \caption{L-mode}
        \label{fig:L-mode_edgepara}
    \end{subfigure}
    \begin{subfigure}{0.45\textwidth}
        \includegraphics[width=\textwidth]{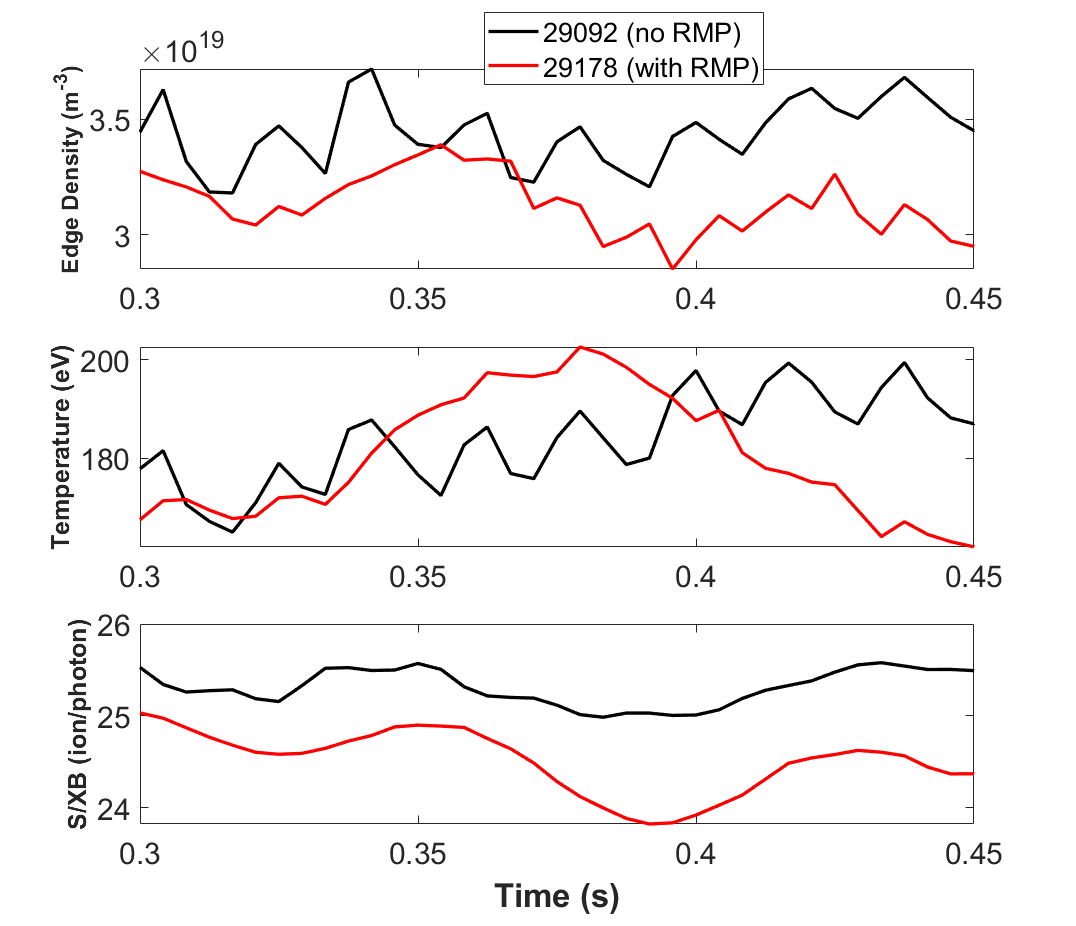}
        \caption{H-mode}
        \label{fig:H-mode_edgepara}
    \end{subfigure} 
    \caption{Time traces of the edge temperature and density and corresponding $S/XB$ coefficients of two L-mode and two H-mode discharges, the axisymmetric discharges are shown in black and RMP discharges are red.}
\end{figure}

The $S/XB$ coefficients are determined from the ADAS database using the local electron temperature and density measured by the Thomson scattering diagnostic at the midplane edge. The precise location for this is important since the $S/XB$ coefficients can vary greatly with changes in temperature and density in this regime. The exact location is determined by a 1-D midplane $D_\alpha$ camera where the radial point of the peak emission is picked and $T_e$ and $n_e$ are interpolated from measurements of the Thomson Scattering diagnostic. The time traces for these parameters are shown in Figure \ref{fig:L-mode_edgepara} for the L-mode discharges and Figure \ref{fig:H-mode_edgepara}. The discharges with RMPs are shown in red. The RMPs do cause the density in the edge to decrease, but generally the temperature is unaffected. This decrease in density leads to insignificant decreases in the $S/XB$. The H-mode discharge presented in Figure \ref{fig:H-mode_edgepara} has more variation through time due to transient ELM events, but the $S/XB$ coefficient shows that the jumps in the edge density and temperature don't have a large impact on the coefficient. Similar to the L-mode scenario, the RMPs do cause the coefficient to decrease slightly.

Using experimentally measured values for all of these terms, the particle confinement time $\tau_p$ can be calculated using a rearranged version of Equation \ref{eqn:one}:

\begin{eqnarray}
    \tau_p = \frac{N}{\Phi_i-\frac{dN}{dt}} 
    \label{eqn:three}
\end{eqnarray}

\begin{figure}
\centering
    \begin{subfigure}{0.42\textwidth}
        \includegraphics[width=\textwidth]{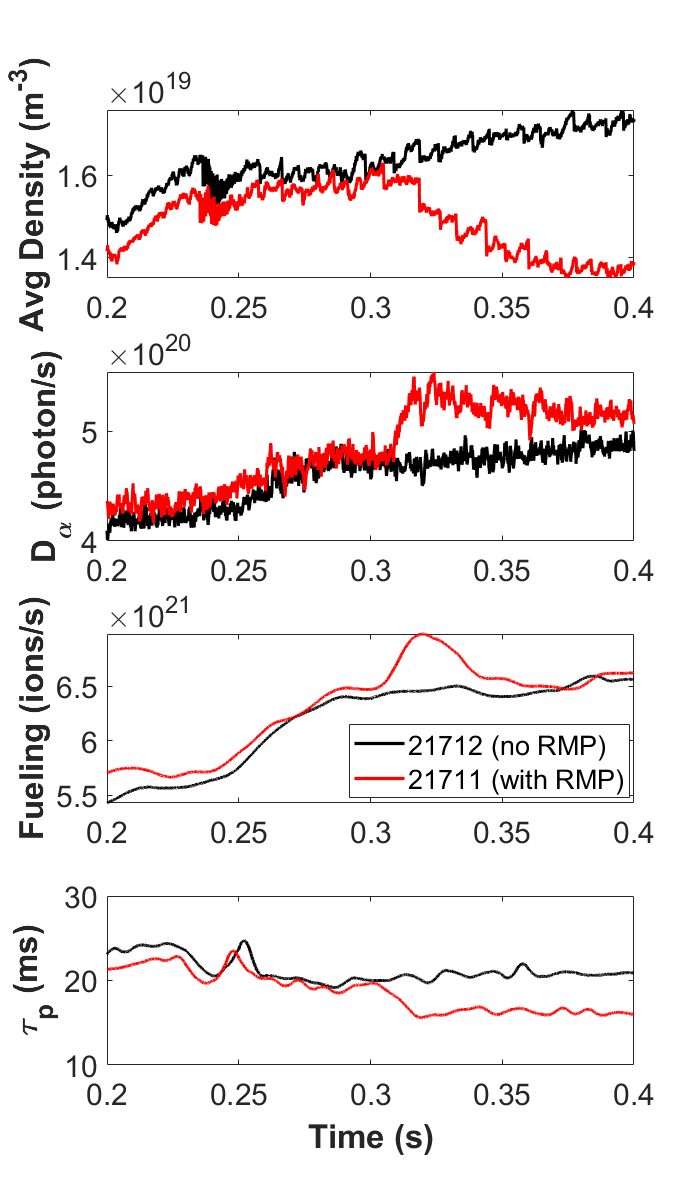}
        \caption{L-mode}
        \label{fig:L-mode_taup}
    \end{subfigure}
    \begin{subfigure}{0.45\textwidth}
        \includegraphics[width=\textwidth]{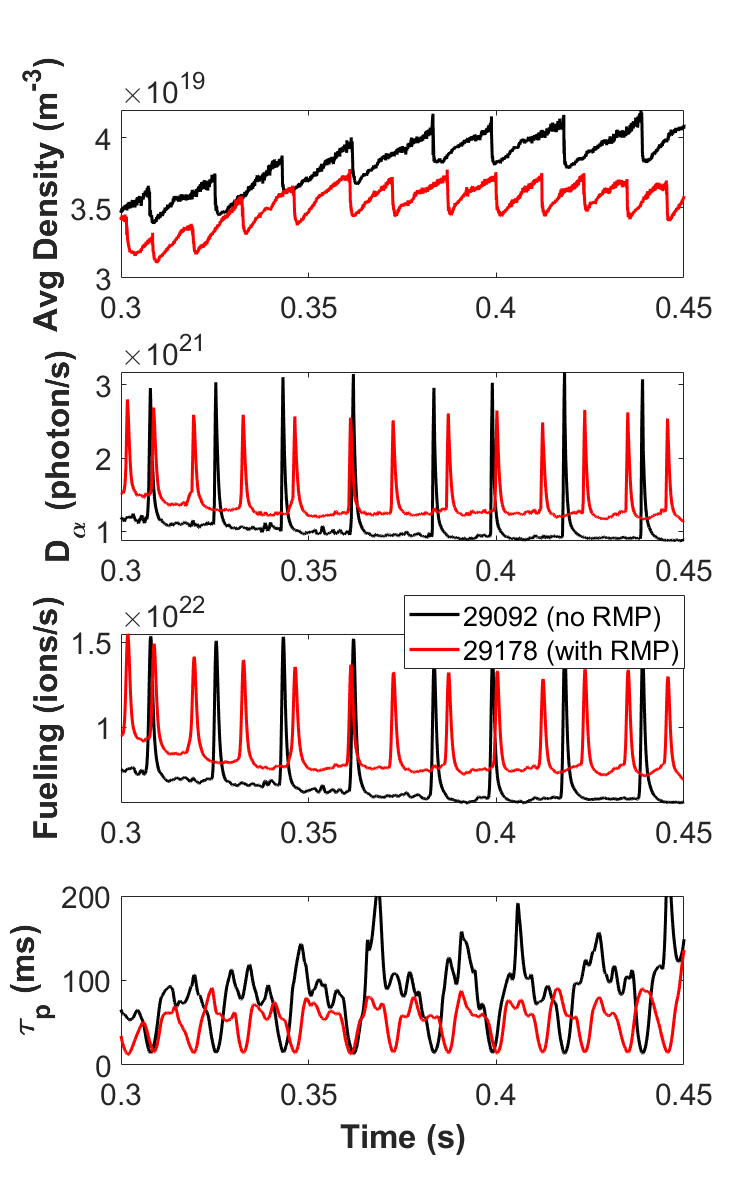}
        \caption{H-mode}
        \label{fig:H-mode_taup}
    \end{subfigure} 
    \caption{Time traces of electron density and $D_\alpha$ emission used in calculating the fueling rate and $\tau_p$ of two L-mode and two H-mode discharges, the axisymmetric discharges are shown in black and RMP discharges are red.}
\label{fig:taup}
\end{figure}

The time traces for the terms in these calculations are shown in Figure \ref{fig:taup}. On the left, the L-mode discharges have the RMP coils turning on just after $0.3s$ which is seen in the drop in density. Simultaneously, the $D_\alpha$ emission increases, and this results in an increase in the calculated fueling rate at that time. Because of the increase in fueling but decrease in density, the confinement time goes down significantly as the RMPs are turned on. This result is approximately a 15\% reduction in the $\tau_p$.  On the right for the H-mode discharges, the RMP coils have already been turned on before the time window, and while the density varies for both cases due to the ELM events, the RMP discharge establishes at a lower density than the axisymmetric case. As with the L-mode scenario, the RMP case again has a higher inter-ELM $D_\alpha$ emission which leads to a higher fueling rate. The average inter-ELM confinement time is also shorter for the RMP case than the axisymmetric, approixmately 10\%. This first estimate indicates that the application of RMPs to L-mode and H-mode discharges could result in a change in both the particle fueling and exhaust (i.e. transport) terms.

\section{\label{sec:multires}Global Multi-Reservoir Particle Balance}
\subsection{MRPB set-up}
To investigate further into how neutral particles play a role in the experimentally observed changes in density with the application of RMPs, especially the particle fueling side, a multi-reservoir particle balance was used. This particle balance was originally composed by Maddison et al \cite{maddison_global_2006} in order to analyze and predict the effectiveness of gas fueling schemes at MAST, but since it accounts for neutral particles and their roles in fueling the plasma, it can also be utilized to track how RMPs may change the neutral components. 

\begin{figure}
    \centering
    \includegraphics[width=\textwidth]{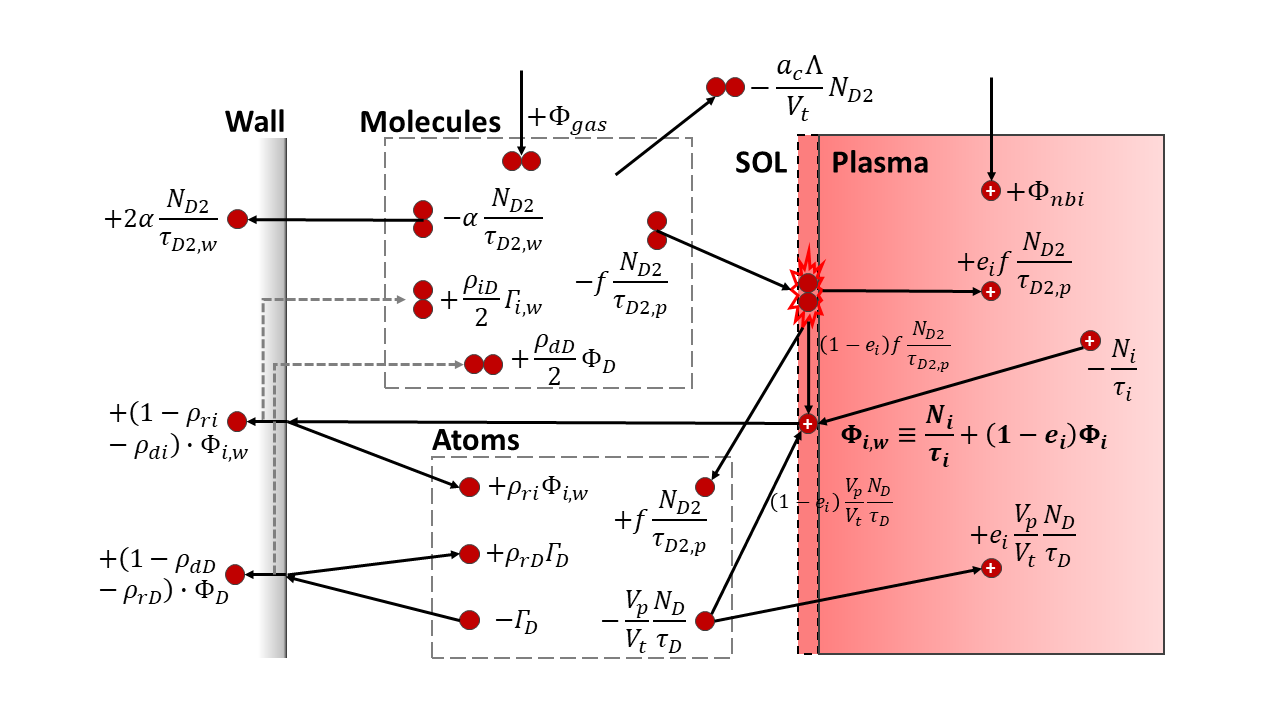}
    \caption{Schematic diagram depicting the 4 reservoirs of the MRPB model and terms from equations \ref{eq:ions}-\ref{eq:atom_flux}.}
    \label{fig:pbal}
\end{figure}

This reservoir balance sets up 4 separate reservoirs inside the vacuum vessel. The plasma, like in the previous section, is counted as one single reservoir, and the other 3 are comprised of each the molecules and the atoms in the vessel and the particles adsorbed to the wall. The following set of equations govern the interactions among the 4 reservoirs:

\begin{eqnarray} \label{eq:ions}
\frac{dN_i}{dt} = -\frac{N_i}{\tau_p} + e_i\Phi_i
\end{eqnarray}

\begin{eqnarray} \label{eq:atoms}
\frac{dN_D}{dt} = -\frac{N_D}{\tau_D} + e_{D2,D}\left(f\frac{N_{D2}}{\tau_{D2,p}}\right) + \rho_{ri}\left(\frac{N_i}{\tau_p}+(1-e_i)\Phi_i\right) + \rho_{rD}\Phi_D
\end{eqnarray}

\begin{eqnarray} \label{eq:molecule}
\frac{dN_{D2}}{dt} = -f\frac{N_{D2}}{\tau_{D2,p}} -\alpha\frac{N_{D2}}{\tau_{D2,w}} + \frac{\rho_{di}}{2}\left(\frac{N_i}{\tau_p}+(1-e_i)\Phi_i\right) + \frac{\rho_{dD}}{2}\Phi_D + \Phi_{gas} - \frac{a_c\Lambda}{V_t}N_{D2}
\end{eqnarray}

\begin{eqnarray} \label{eq:wall}
\frac{dN_w}{dt} = \left(1-\rho_{ri}-\rho_{di}\right) \left(\frac{N_i}{\tau_p}+\left(1-e_i\right)\Phi_i\right) + \left(1-\rho_{rD}-\rho_{dD}\right)\Phi_D + 2\alpha\frac{N_{D2}}{\tau_{D2,w}}
\end{eqnarray}

where $\Phi_i$ and $\Phi_D$ are

\begin{eqnarray} \label{eq:ion_flux}
 \Phi_i = e_{D2,i}\left(f\frac{N_{D2}}{\tau_{D2,p}}\right) + \frac{V_p}{V_t}\frac{N_D}{\tau_D}   
\end{eqnarray}

\begin{eqnarray} \label{eq:atom_flux}
\Phi_D = \left(1-\frac{V_p}{V_t}\right)\frac{N_D}{\tau_D}    
\end{eqnarray}

In this notation, the total particles in each reservoir is represented as $N_i$, $N_D$, $N_{D2}$, and $N_w$ for the ions, atoms, molecules, and wall inventory respectively. Particles remain in each reservoir for some average dwell time represented by each $\tau$ term, except for the wall where particles do not reenter the other reservoirs in this setup. For a full treatment of every term in this particle balance, please refer to Maddison et al \cite{maddison_global_2006}. The following will describe modifications made to their original formulation based on observations from experiment and modeling.

\subsection{Modifying MRPB interaction terms}
The terms $\rho_{ri}$ and $\rho_{di}$ refer to the percent of ions impacting a divertor or wall surface that either reflect as an atom or desorb as half of a molecule respectively. Correspondingly, the terms $\rho_{rD}$ and $\rho_{dD}$ refer to the percent of atoms impacting the surface that either reflect back as an atom or desorb as half of a molecule. From this, $(1-\rho_{ri}-\rho_{di})$ is the portion of ions lost to the wall reservoir, and $(1-\rho_{rD}-\rho_{dD})$ is the portion of atoms lost. Originally, the estimated values were $\rho_{ri}=\rho_{di}=\rho_{rD}=0.5$, and the last value, $\rho_{dD}$, could be varied to allow the model to better fit experimental data. However, simulation results from the 3-D edge modeling code EMC3-EIRENE, a description of which will be given in Section \ref{sec:EMC3}, were able to better constrain each of the values. The EMC3-EIRENE code calculates each incoming ion's or atom's impact angle and energy and uses the TRIM database to find the likelihood of the particle to reflect as an atom or desorb as a molecule. By tracing individual particles with EMC3-EIRENE and tabulating whether each wall impact resulted in an atom or molecule, the average probabilities for reflection were calculated to be $\rho_{ri}\approx0.23$ and $\rho_{rD}\approx0.85$. This was consistent across fueling levels for both L-mode and H-mode scenarios, varying by no more than 2\%. Therefore, the following coefficients were used in the MRPB model: $\rho_{ri}=0.23$, $\rho_{di}=0.77$, $\rho_{rD}=0.85$, and $\rho_{dD}=0.15$. Also, instead of adjusting only $\rho_{dD}$ to better fit the experimental data, all coefficients were lowered equally to account for the wall-pumping, which tended to lead to more accurate model results than the previous methods. These modifications can be seen summarized in Table \ref{tab:surface_terms}

\begin{table}
    \centering
    \begin{tabular}{|c|c|c|c|c|c|}
    \hline
    \textbf{Parameter} & $\rho_{ri}$ & $\rho_{di}$ & $\rho_{rD}$ & $\rho_{dD}$ & Free parameter\\
    \hline \hline
    Original & 0.5 & 0.5 & 0.5 & 0.5 & $\rho_{dD}$\\
    \hline
    Modified & 0.23 & 0.77 & 0.85 & 0.15 & all equally\\
    \hline
    \end{tabular}
    \caption{Original and modified surface interaction terms for ion ($\rho_{ri}$ and $\rho_{di}$) and atomic deuterium ($\rho_{rD}$ and $\rho_{dD}$)}
    \label{tab:surface_terms}
\end{table}

Additionally, particle tracing from EMC3-EIRENE indicated the MRPB equations required a modification for its treatment of molecular dissociation. Originally, the molecular flux impacting the plasma, $f\frac{N_{D2}}{\tau_{D2,p}}$, would result in one ion and and one atom each. This caused molecular dissociation to account for approximately 56\% of the fueling ions. However, the particle tracing from EMC3-EIRENE simulations showed that ions born from molecular dissociation events, or promptly after a molecular dissociation, made up approximately 24\% of the ions fueling the plasma for both L-mode and H-mode. To adjust the MRPB to more accurately capture the processes in the SOL, a pair of coefficients $e_{D2,D}$ and $e_{D2,i}$ were introduced to allow a fraction of molecular dissociation events to result in 2 atoms instead of 1 atom and 1 ion. These were set such that $e_{D2,D}+e_{D2,i}=2$. In order constrain the MRPB to 24\% of the fueling ionizations from molecules, $e_{D2,D}$ was required to be about 1.28, so that 64\% of the molecular flux became 2 atoms and 36\% became 1 atom and 1 ion.

\subsection{Introducing a fueling efficiency}
Initially, as written in equation \ref{eqn:one}, it is assumed that all ionizations calculated from this method of using the $D_\alpha$ emission will fuel the plasma. However, this may not be the case as some of the photons collected by the filterscope may be from ionizations occurring in the SOL that would instead promptly stream along open field lines to the divertor and never fuel the plasma. Therefore, a fueling efficiency should be introduced, which is the percent of ionizations that result in an ion which fuels the confined plasma. This makes equation \ref{eqn:one}, the single reservoir balance, also match equation \ref{eq:ions}, the plasma reservoir from the MRPB. For the MRPB equation, however, the $\tau_p$ values come from the calculations in section \ref{sec:partconfine} and $\Phi_i$ is calculated internally.

Results from the MRPB then become an iterative process where there are essentially 2 knobs to turn to allow for best fit between the experimental results and MRPB. These are: $e_i$, the fueling efficiency set the same in both equations \ref{eqn:one} and \ref{eq:ions}; and $\rho_{ri}$, $\rho_{di}$, $\rho_{rD}$, and $\rho_{dD}$, the wall reflection and adsorption terms which are all lowered equally. Three parameters are sampled from the MRPB to compare to the experimental results: electron density, from the total ions and plasma volume; neutral pressure, from the total molecules and neutral volume; and $D_\alpha$ emission, from the ionization rate. The time varying values for these quantities can then be compared to the values measured directly in the experiment. This process is implemented using the Matlab ODE45 function.

\subsection{MRPB results}

\begin{figure}
\centering
    \begin{subfigure}{0.45\textwidth}
        \includegraphics[width=\textwidth]{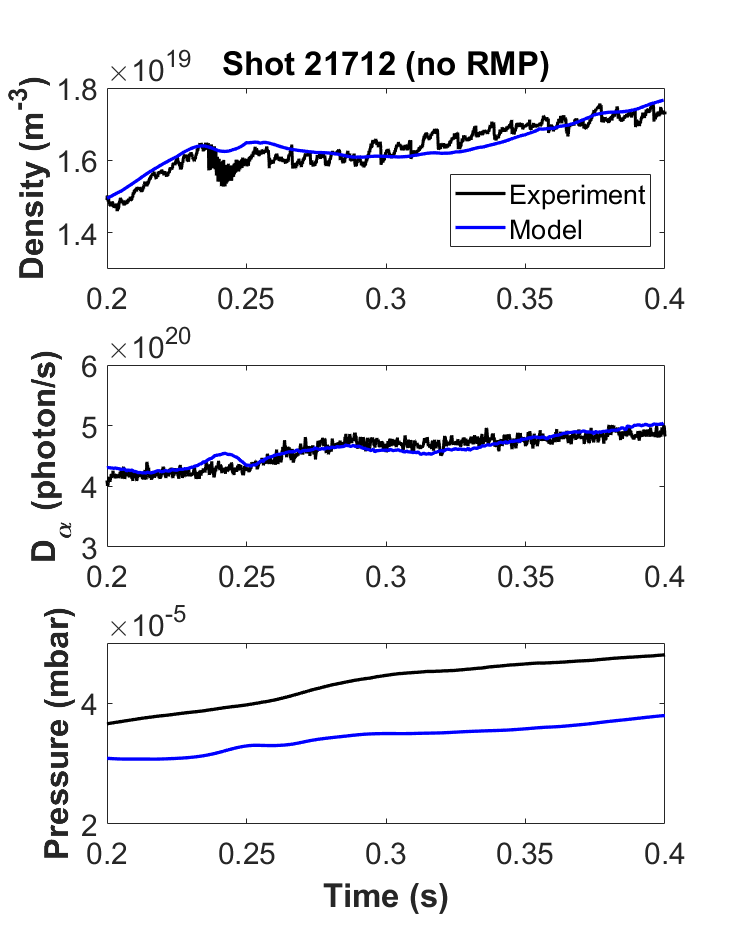}
    \end{subfigure}
    \begin{subfigure}{0.45\textwidth}
        \includegraphics[width=\textwidth]{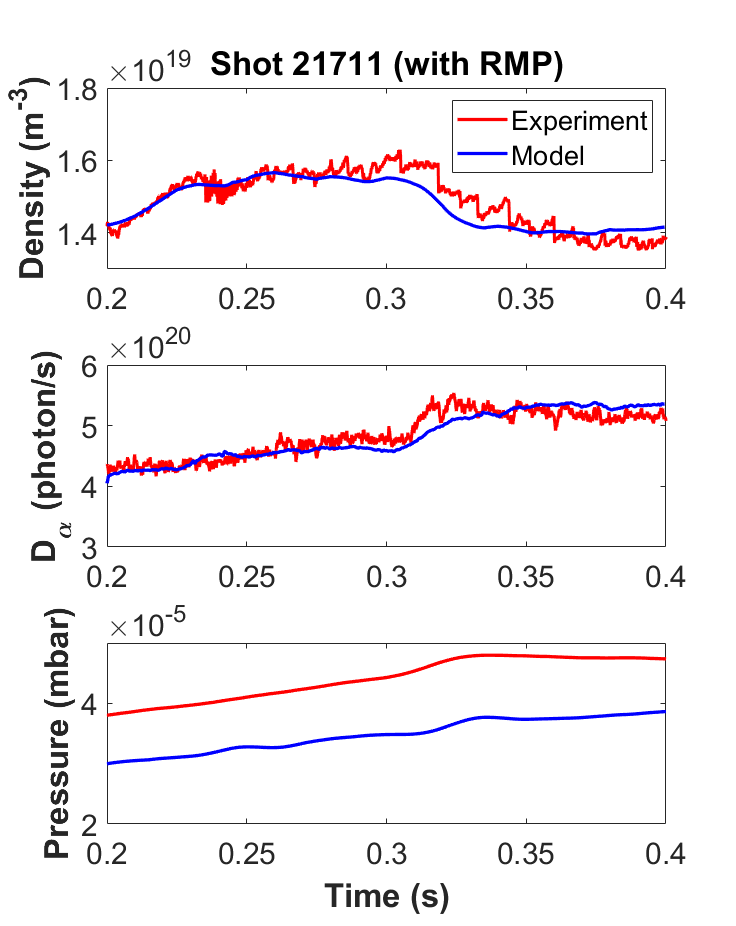}
    \end{subfigure} 
    \caption{Time evolution of 2 L-mode discharges, non-RMP case on the left and RMP case on the right, with the experimental measurements in black and red and the MRPB calculated values in blue for both.}
\label{fig:L-mode_MRPB}
\end{figure}

The MRPB model is able to reproduce the experimental values quite well, with the calculated $\tau_p$ values and the hand-tuned values of $e_i$, $\rho_{ri}$, $\rho_{di}$, $\rho_{rD}$, and $\rho_{dD}$ as inputs. The L-mode discharges are shown in Figure \ref{fig:L-mode_MRPB} with both the axisymmetric case (left) and RMP case (right) where the MRPB calculations are depicted in blue. For the L-mode cases, the calculated density and photon emission matched the experimental values very well. The MRPB was able to capture the density pump-out and coinciding increase in $D_\alpha$ emission as the RMPs were turned on just after 0.3s. It doesn't capture the fast timescale of the small density and $D_\alpha$ fluctuations due to some necessary smoothing in the calculation of $\frac{dN_i}{dt}$ in order to obtain the $\tau_p$, but those timescales are outside the scope of this analysis. The vessel neutral pressure is slightly underestimated, but well within an order of magnitude, compared to the experiment, but direct comparison with this value was not expected to match as well since there was only one neutral pressure gauge in the device. It was located at the outboard midplane at one toroidal point, which is not near the fast dynamics of the SOL and divertor regions, therefore any changes in those regions would take vacuum neutral timescales to reach the gauge. This is seen more clearly in the H-mode discharges. For the L-mode RMP case, the model does recreate the small increase in neutral pressure after the RMPs are turned on as particles are expelled from the plasma, and for both cases, it denotes the slow increase in pressure through the time of the discharge.

\begin{figure}
\centering
    \begin{subfigure}{0.45\textwidth}
        \includegraphics[width=\textwidth]{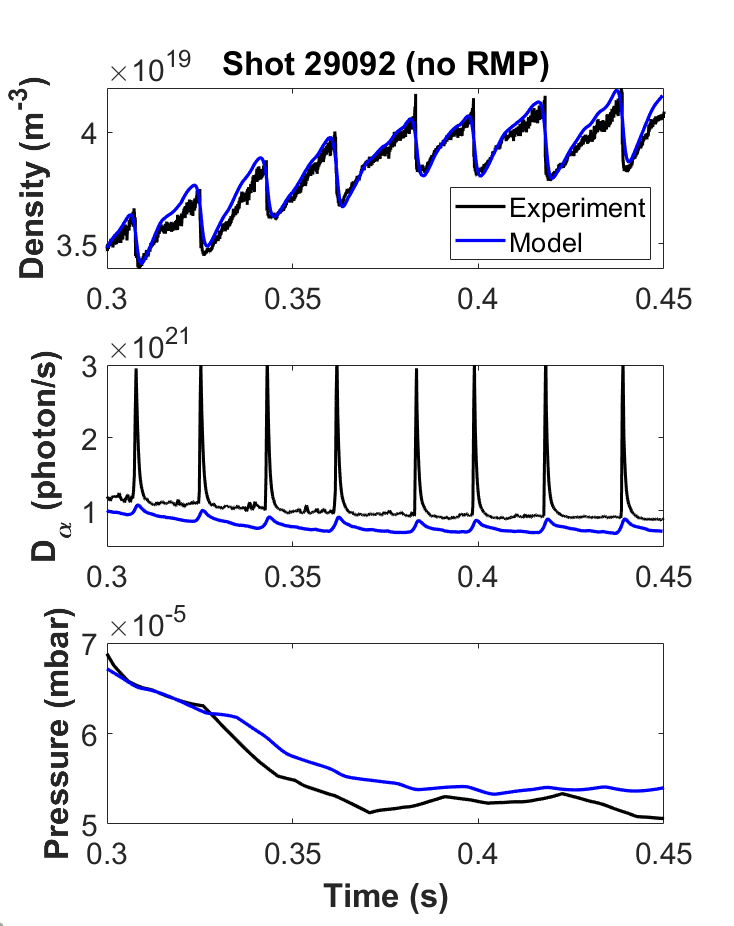}
    \end{subfigure}
    \begin{subfigure}{0.45\textwidth}
        \includegraphics[width=\textwidth]{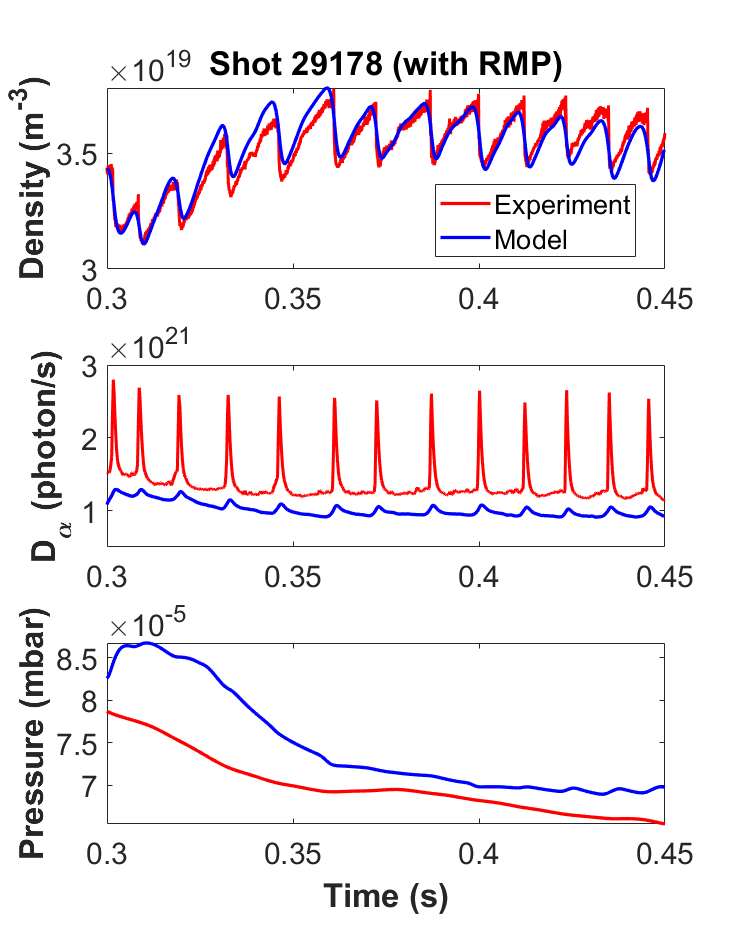}
    \end{subfigure} 
    \caption{Time evolution of 2 H-mode discharges, non-RMP case on the left and RMP case on the right, with the experimental measurements in black and red and the MRPB calculated values in blue for both.}
\label{fig:H-mode_MRPB}
\end{figure}

The results from the MRPB for the H-mode cases are shown in Figure \ref{fig:H-mode_MRPB}. Once again, the MRPB model is able to, with moderate accuracy, capture the general dynamics of the H-mode discharges and the changes created by the RMP application. The calculated density, including the large drops caused by the ELMs and reduction due to the RMPs, matches very well with the experiment. It does under estimate the inter-ELM $D_\alpha$ emission but still shows the increased baseline for the RMP discharge compared to the axisymmetric case. The large spikes in $D_\alpha$ emission are not recreated in the model, due once again to the smoothing necessary in the $\frac{dN_i}{dt}$ calculations for the $\tau_p$. This is allowable for this analysis, though, because we are not attempting to look at the ELM events themselves or on their timescales and the model treats them similarly between the two cases in the end. For the H-mode cases, the total neutral pressure matches more closely in absolute values to the experimental ones compared to the L-mode discharges. However, for the RMP case, the MRPB predicts a large increase in neutral pressure as the RMPs begin to take effect. This is due to the sudden drop in plasma density, which results in a large number of particles entering the neutral population, building up the pressure before they can be pumped away by the vessel pumps. This is not captured in the experimental values which could be due the location of the pressure gauge and the time response of it as well.

\subsection{Modifying $e_i$ vs. $\tau_p$}
Originally, the assumption is made that any change caused by the RMPs to the density and ionizations would result in a change in confinement time $\tau_p$ while leaving the fueling efficiency $e_i$ constant and equal between axisymmetric and RMP scenarios. This may not be the case however, and discerning whether the change is in $\tau_p$, essentially a transport effect, or not is the basis for this study. Instead of using a constant $e_i$ for both axisymmetric and RMP, the calculation of $\tau_p$ and the MRPB can use a modified value that varies when the RMPs are turned on. In this alternative method, the value for $e_i$ is reduced by 15\% when the RMPs are applied in the L-mode scenario and by 10\% for the entire H-mode scenario as the RMPs are applied before the window used for analysis. The change from an RMP-induced reduction in $\tau_p$ to a reduction in fueling efficiency results in identical time traces from the MRPB.

This is not altogether intuitive, but by looking at the coupled differential equations used in the model, one can understand how this occurs. Because the neutral population starts the same for both scenarios, the number of neutral particles that become ionized is also the same. For the case with lower confinement time, when the RMPs are turned on, the same number of ions reach the bulk plasma, but more particles are exhausted from the plasma, leading to a larger flux on the wall. On the other hand, for the case with lower fueling efficiency, the same number of neutral particles are ionized, but some never make to the bulk plasma. Instead, those flow directly to the wall. In either case, this leads to less particles in the plasma and more particles hitting the wall, and because the same percentage is used, this has the exact same effect.

\section{\label{sec:EMC3}EMC3-EIRENE}

The EMC3-EIRENE code is a coupled Monte Carlo 3-D edge plasma and neutral boundary numerical modeling tool. The EMC3 code is used to simulate the plasma edge and scrape-off layer by solving a plasma fluid model with particle, energy, and momentum sources. The Braginskii equations of the plasma model in EMC3 updates the particle sourcing from EIRENE and interated until convergence for a self-consistent solution of the plasma edge between ions and neutral particles. For a more full description see \cite{feng_3d_1999} and \cite{frerichs_volumetric_2021}. For this modeling work, multiple simulations for both H-mode and L-mode were conducted, but for each, the transport terms $D_\perp$ and $\chi_\perp$ were spatially fixed coefficients and held constant between differing particle fueling schemes in order to better understand the effect of the changes of the magnetic field on plasma conditions. The total ionization rate in the simulation domain was then changed to best match the modeling conditions to the experimental measurements. When operating with a fixed ionization rate, particles are sourced from the NBI and gas puffing, but EMC3-EIRENE imposes a global rescaling factor such that the total ionization rate in the plasma domain is exactly what was specified in the simulation input.

\begin{figure}
    \centering
    \includegraphics[width=0.8\textwidth]{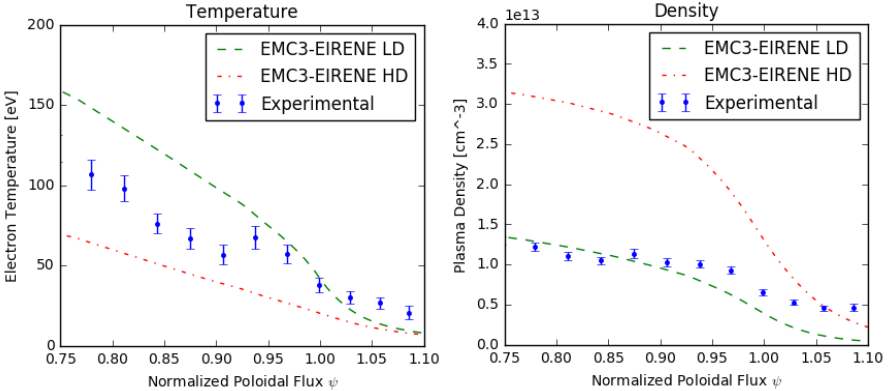}
    \caption{Comparison of electron temperature and density from EMC3-EIRENE modeling at two different fueling rates to experimental data in blue for an axisymmetric L-mode discharge.}
    \label{fig:EMC3-Lmode}
\end{figure}

The L-mode cases used $D_\perp=1.15\, \frac{m^2}{s}$ and $\chi_\perp=4\, \frac{m^2}{s}$ for its simulations.  The axisymmetric case for the L-mode discharge is shown in Figure \ref{fig:EMC3-Lmode} with the mid-plane electron temperature and density plotted against the normalized poloidal flux coordinate for the edge. Here the green profiles represent the low fueling (low density) simulation and the red represents the high fueling (high density). The blue markers come from the mid-plane Thomson scattering diagnostic. Since the low density case best matches the experimental profile, this case will be used to make comparisons to the experiment and used for further analysis.

\begin{figure}
    \centering
    \includegraphics[width=0.8\textwidth]{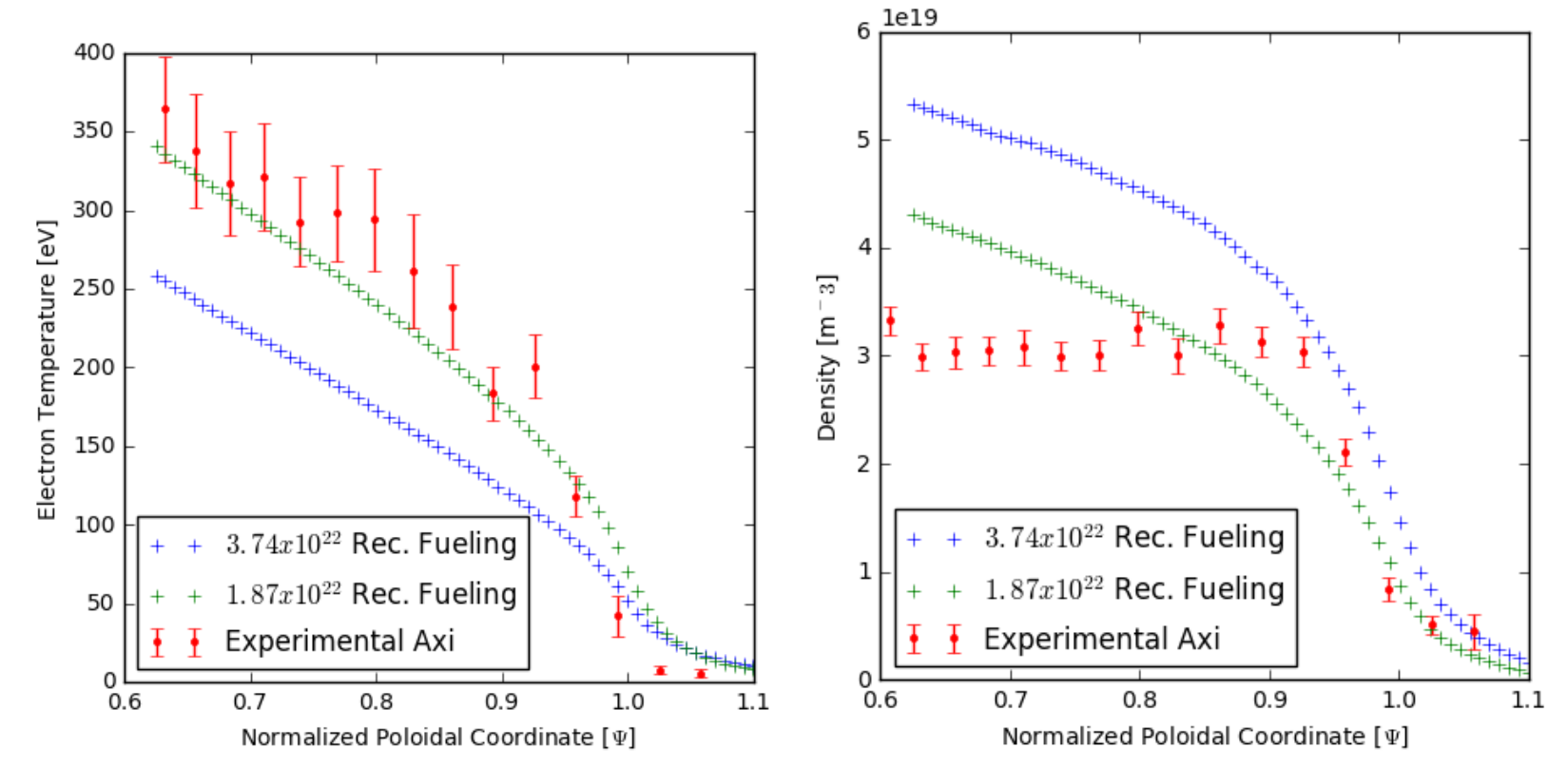}
    \caption{Comparison of electron temperature and density from EMC3-EIRENE modeling at two different fueling rates to experimental data for an axisymmetric H-mode discharge.}
    \label{fig:EMC3-Hmode}
\end{figure}

For the H-mode cases, the NBI was operating, so to account for that a particle flux of $3.74\times10^{20}$ ion/s is sourced from the inner simulation boundary, and $P_{SOL}=1.2$ MW split evenly between the ions and electrons. The transport terms are set to $D_\perp=0.3 \, \frac{m^2}{s}$ and $\chi_\perp=3.0\, \frac{m^2}{s}$. The fueling scan for these cases are shown in Figure \ref{fig:EMC3-Hmode} with the mid-plane temperature and the density plotted again on the normalized flux coordinate. The blue profile represents the low fueling (low density) case and the green the high fueling (high density) case. The experimental profiles in Figure \ref{fig:EMC3-Hmode} are shown in red from the Thomson scattering measured during inter-ELM periods. For the H-mode cases, the high fueling (high density) case will be used for analysis as it matches the experimental profiles best. For these cases, the temperature matches quite well, but the density profile does not follow as well the flattening in the core compared to experiment. This is most likely due to the fact that the transport terms were not varied radially. While the perpendicular transport may change radially through the plasma, especially in the pedestal, changing the transport terms in order to match the experimental data better introduces a second fitting parameter to try to match the EMC3-EIRENE results to the experimentally measured values. This would also cause the focus to be on correctly picking the radially varying transport terms and take away from the emphasis of this study, which is to analyze whether fueling is impacted by the RMPs, while transport is assumed to be constant. 

\subsection{Impact of magnetic field structure}

For the axisymmetric cases, the magnetic field structure used as an input for EMC3-EIRENE comes directly from the experiment's EFIT calculations. This is typical for modeling experiments already conducted \cite{feng_3d_1999}. However, the RMP cases are more complicated to model. Because the edge region is so vital to the particle fueling and exhaust of the plasma, a high fidelity treatment is needed to determine the magnetic field structure. To do this, MARS-F was used to model the plasma response to the application of the RMPs. MARS-F is a single-fluid, resistive MHD model that uses inputs from measured plasma profiles, plasma rotation, and an EFIT equilibrium to model the magnetic field due to the plasma response \cite{liu_toroidal_2013}. This modeled magnetic field tends to match the lobes near the X-point and strike splitting on the divertor well \cite{kirk_understanding_2013}. 

\begin{figure}
\centering
    \begin{subfigure}{0.4\textwidth}
        \includegraphics[width=\textwidth]{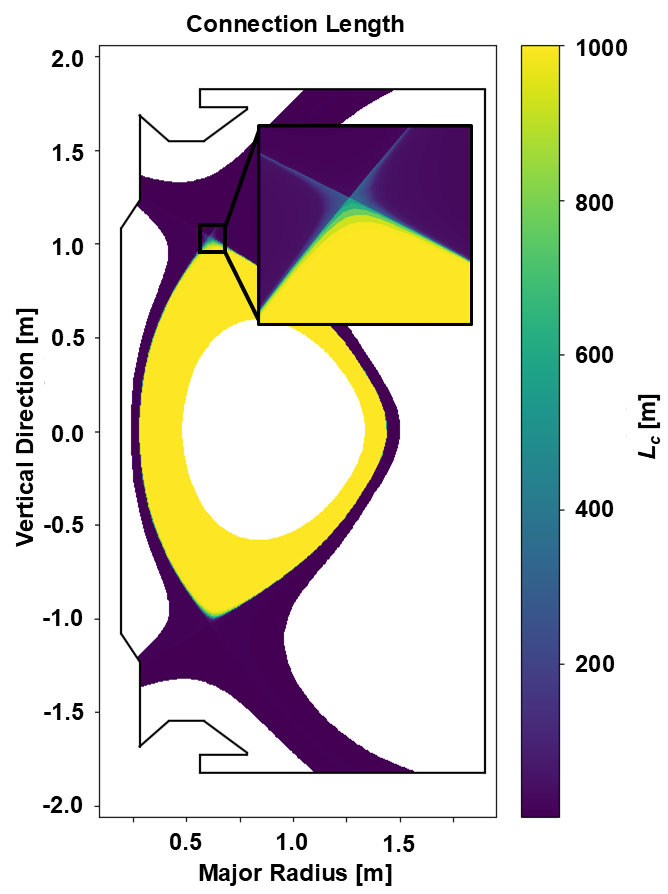}
    \end{subfigure}
    \begin{subfigure}{0.4\textwidth}
        \includegraphics[width=\textwidth]{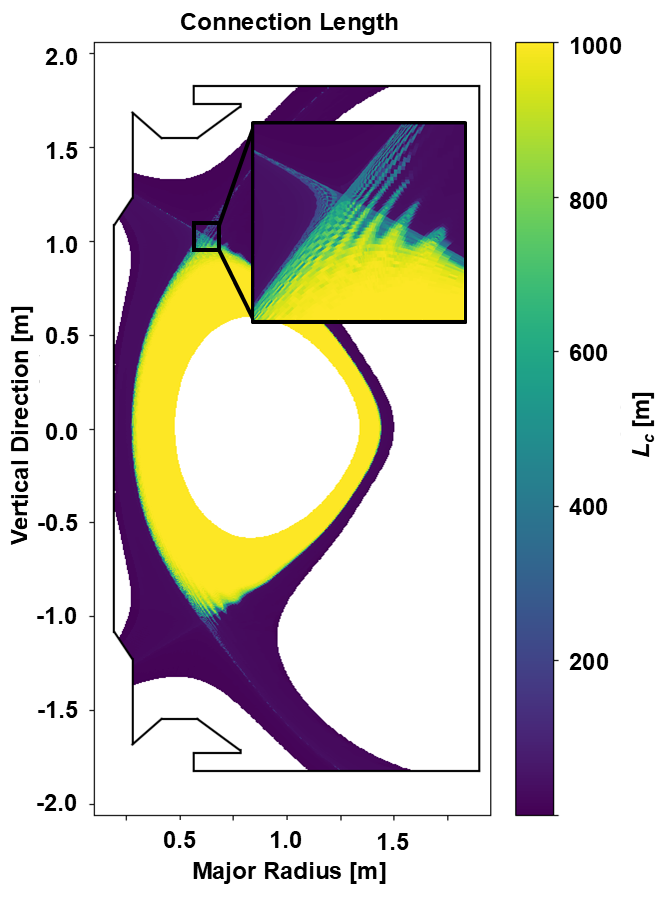}
    \end{subfigure} 
    \caption{Magnetic connection length, non-RMP case on the left and RMP case on the right, plotted for the CDN L-mode on an R-Z slice.}
\label{fig:L-mode_Lc}
\end{figure}

\begin{figure}
\centering
    \begin{subfigure}{0.4\textwidth}
        \includegraphics[width=\textwidth]{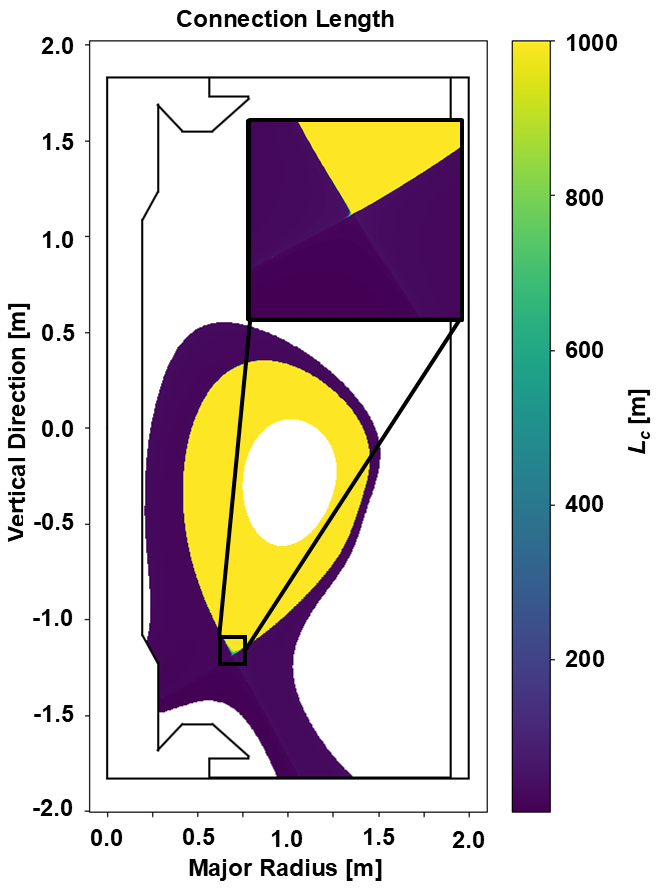}
    \end{subfigure}
    \begin{subfigure}{0.4\textwidth}
        \includegraphics[width=\textwidth]{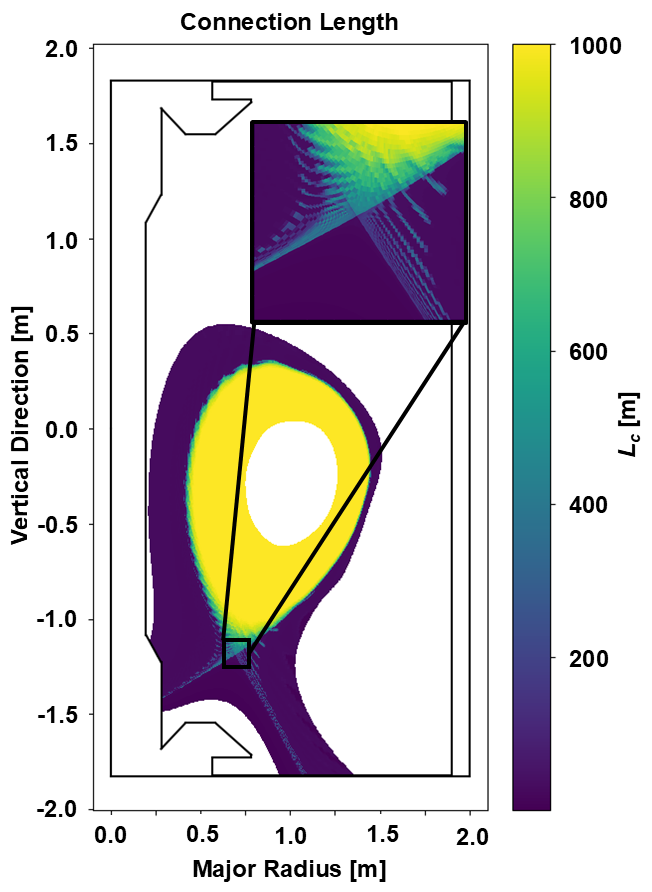}
    \end{subfigure} 
    \caption{Magnetic connection length, non-RMP case on the left and RMP case on the right, plotted for the LSN H-mode on an R-Z slice.}
\label{fig:H-mode_Lc}
\end{figure}

To begin looking at how RMPs may alter the neutral fueling and exhaust of the plasma, we look at how the RMPs change the magnetic structure of the plasma. Because parallel transport is orders of magnitude higher than perpendicular transport, plasma particles on open field lines that connect to physical surfaces will be exhausted at a much higher rates than those on closed field lines. It is well known that RMPs modify the magnetic field structure in the edge of the plasma, causing the edge region to become chaotic and areas of the plasma that were previously closed field lines with good flux surfaces to now be magnetically connected to the divertor surface. To visualize this change, connection length plots for axisymmetric case and RMP case for the L-mode discharge can be seen in Figure \ref{fig:L-mode_Lc}. This samples every node of the simulation domain in an RZ slice and traces forwards and backwards along the magnetic field until it either intersects a surface or hits the computational limit in path-length. For these simulations, a limit of 1000m was used as the path-length limit. For the axisymmetric case, on the left in Figure \ref{fig:H-mode_Lc}, the plot has closed field lines depicted with yellow for the confined region of the plasma. Moving out of the region, across the separatrix, the SOL has very short connection lengths, which are depicted with purple. For this case, the confined region is bounded by a discrete and continuous surface that is the separatrix. For the RMP case, the right plot in Figure \ref{fig:L-mode_Lc}, the chaotic region appears here in green and blue between the closed field lines, in yellow, and short field lines, in purple. The chaotic region is largest near the X-points and thin near the outer edge, but as will be shown in the next section, this thin region has an out-sized impact on particle fueling. The medium connection length field lines indicate that there are now regions of the previously well confined plasma that are magnetically connected to vessel surfaces. 

The same connection length plots for the H-mode scenarios can be seen in Figure \ref{fig:H-mode_Lc}. As with the L-mode plots, the yellow regions indicate closed field lines and the purple regions indicate short connection lengths. The H-mode scenarios were in a lower single null configuration, which is why there is now only a lower X-point and divertor regions. For the axisymmetric case in Figure \ref{fig:H-mode_Lc}, similar to the L-mode one, there is a continuous surface separating the confined region from the SOL. When the RMPs are added, on the right in Figure \ref{fig:H-mode_Lc}, there is again a thin chaotic region with medium connection length field lines separating the confined region and the traditional SOL region. For this case, the perturbed region is larger and reaches deeper into the previously well confined area of the plasma.

\subsection{Quantifying fueling for axisymmetric and RMP discharges}

\begin{figure}
\centering
    \begin{subfigure}{0.4\textwidth}
        \includegraphics[width=\textwidth]{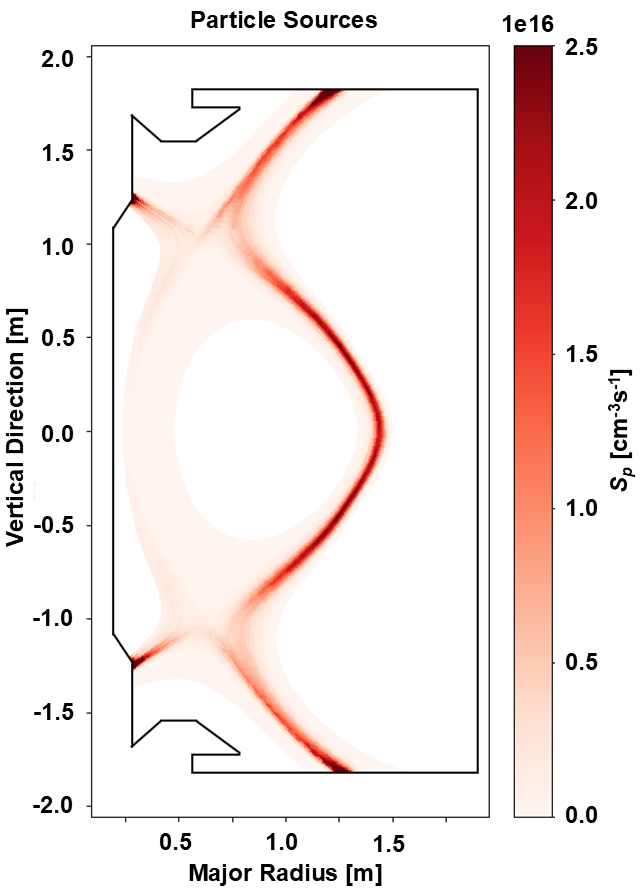}
        \label{fig:L-mode_noRMP_Sp}
    \end{subfigure}
    \begin{subfigure}{0.4\textwidth}
        \includegraphics[width=\textwidth]{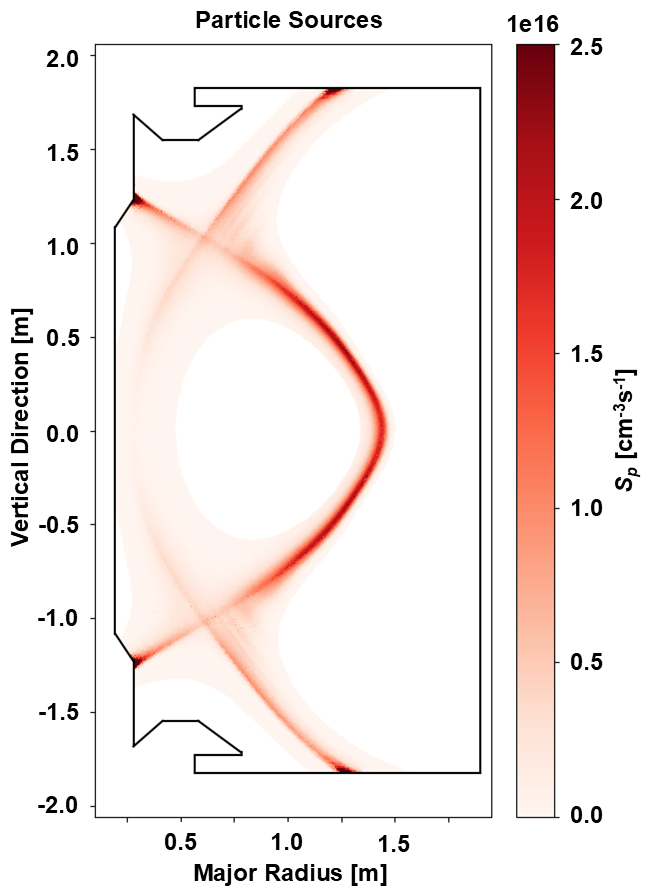}
        \label{fig:L-mode_RMP_Sp}
    \end{subfigure} 
    \caption{Total volumetric ionization rate, or neutral fueling, for the L-mode non-RMP case on the left and RMP case on the right.}
\label{fig:L-mode_Sp}
\end{figure}

\begin{figure}
\centering
    \begin{subfigure}{0.4\textwidth}
        \includegraphics[width=\textwidth]{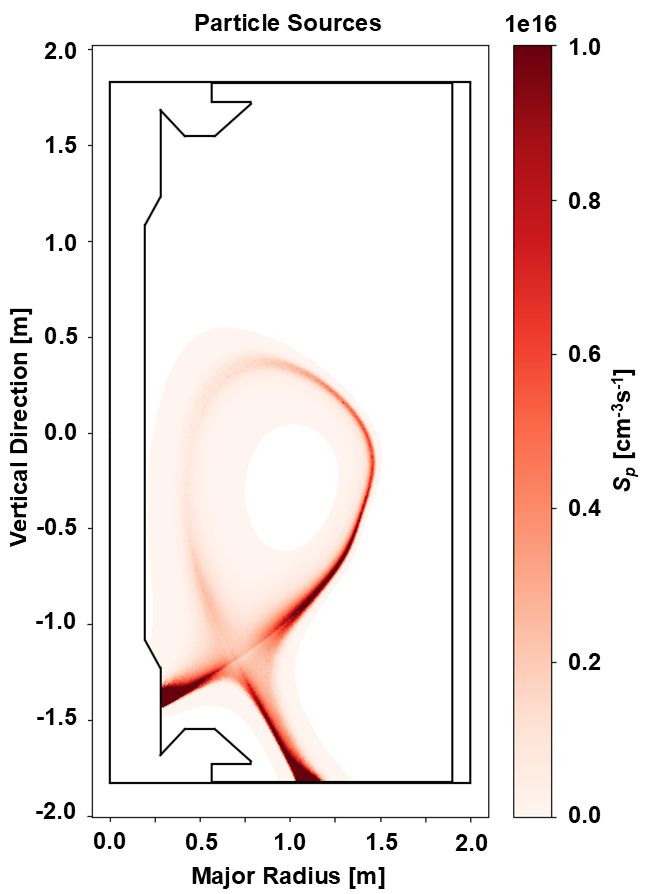}
    \end{subfigure}
    \begin{subfigure}{0.4\textwidth}
        \includegraphics[width=\textwidth]{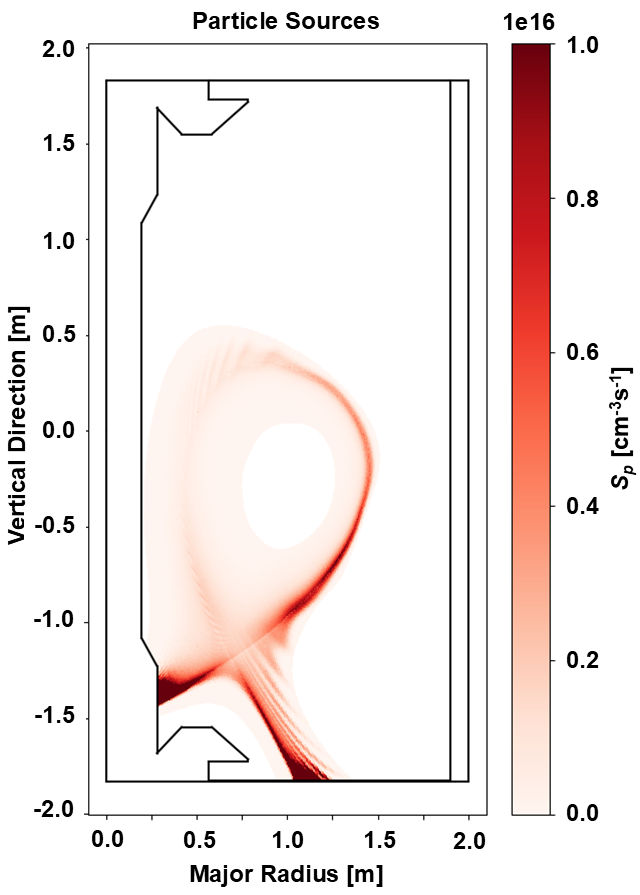}
    \end{subfigure} 
    \caption{Total volumetric ionization rate, or neutral fueling, for the H-mode non-RMP case on the left and RMP case on the right.}
\label{fig:H-mode_Sp}
\end{figure}

The plasma density, as has been described previously, is balanced by both transport out of the plasma and neutral particle fueling into the plasma. Neutral fueling to the plasma occurs via ionizations of particles either recycling from the divertor or being puffed in from neutral injections. In MAST, the distinction between these two sources is mostly neglegible. This is because of the open divertor and vessel design used. Many ions that impact the divertor surfaces recombine at the surface, recycle as neutrals into the divertor legs, and are promptly reionized. These ions, however, tend to flow directly back into the divertor surface and do not fuel the confined plasma. This can be seen in the large ionization rates in Figures \ref{fig:L-mode_Sp} and \ref{fig:H-mode_Sp} near the divertor surfaces. Because of the open structure of the divertors, a significant number of neutrals reflect away from the divertor legs and then are able to ballistically travel throughout the vessel before interacting with the plasma edge on the outboard side \cite{casali_neutral_2020}. Gas puffed from the outer vessel ports also travels in this manner so that recycled neutrals and puffed neutrals become mixed and both fuel the plasma similarly.

Volumetric ionization rates, as calculated with EMC3-EIRENE, for the L-mode and H-mode scenarios have been plotted in Figure \ref{fig:L-mode_Sp} and Figure \ref{fig:H-mode_Sp} respectively. All of the cases have very similar properties where there is a large number of ionizations occurring near the divertor targets and also the edge of the plasma on the outboard side. These higher rates of ionization are indicated with darker red coloration. The RMP cases have visible lobe structures near the X-point, and those have been observed in experiment \cite{harrison_characteristics_2014}. Most of the ionization events for both cases, however, do not fuel the plasma because the ions are created on open field lines where parallel transport mechanisms dominate and tend to flow directly towards a divertor before having the chance to migrate inward. Even in the outboard edge, those ions created on open field lines will primarily flow towards the divertor. As has been shown already, the application of RMPs changes the magnetic structure in the edge and where the open field lines lie, so these RMPs may change the particle fueling with regards to the magnetically confined region. Quantifying the particle fueling that occurs in the confined region and comparing them between the axisymmetric case and RMP case can then allow us to determine the relative fueling efficiency between the two.

Before we can quantify the particle fueling to the confined region, first we must define what the confined region is. At first glance, one could expect it to be the yellow regions in Figures \ref{fig:L-mode_Lc} and \ref{fig:H-mode_Lc}. For the axisymmetric cases, that would be accurate as the confined region is defined by the area inside the separatrix that has closed field lines with infinite connection length, and everything outside the separatrix is not part of the confined region. For the RMPs cases, however, the yellow region is determined by the area of the simulation that is not magnetically connected to the divertor targets \textit{up to the computational limit}. If that computational limit is changed, then the region will change. Indeed, when the limit is set to follow field lines only to 200m, the volume of the confined region becomes larger, and when the limit is set to 5000m, open field lines are traced even deeper into the plasma and the confined region becomes smaller. Changing the volume of the confined region then changes the integrated fueling to that region. One could continue to trace magnetic field lines for progressively longer limits in order to find the true edge of the confined region for the RMP cases, however, eventually those field lines become long enough that cross field transport mechanisms will begin to compete with parallel transport. Therefore, the concept of fueling to the confined region cannot be based solely on open or closed field lines for the RMP cases with an arbitrarily imposed field line following cutoff.

Because EMC3-EIRENE is a fully 3-D model, it is possible to analyze important quantities as profiles along field lines as they precess poloidally and toroidally around the plasma. This allows us to characterize confined and open regions not just by the magnetic connection length but by the plasma behavior along those flux tubes. For this analysis, we look at two particular quantities: particle source rate and plasma flow speed. The particle source rate is important because it is the quantity of interest for understanding how the fueling changes. The plasma flow speed is important because this will determine whether a specific region of plasma will behave like the SOL with flow directly towards a divertor target or like the confined region with flow away from a divertor target and along the field line numerous times around the tokamak. 

\begin{figure}
\centering
    \begin{subfigure}{0.45\textwidth}
        \includegraphics[width=\textwidth]{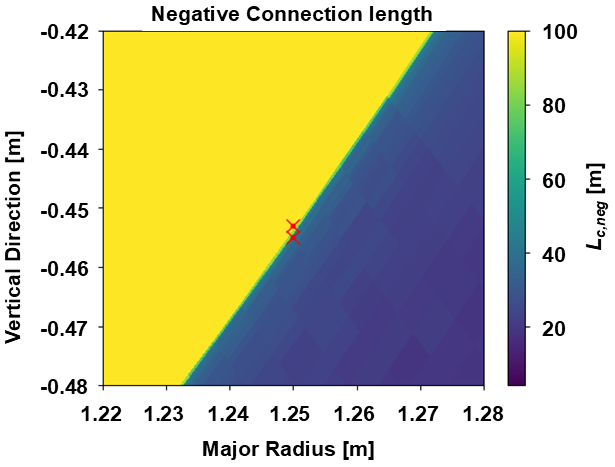}
    \end{subfigure}
    \begin{subfigure}{0.45\textwidth}
        \includegraphics[width=\textwidth]{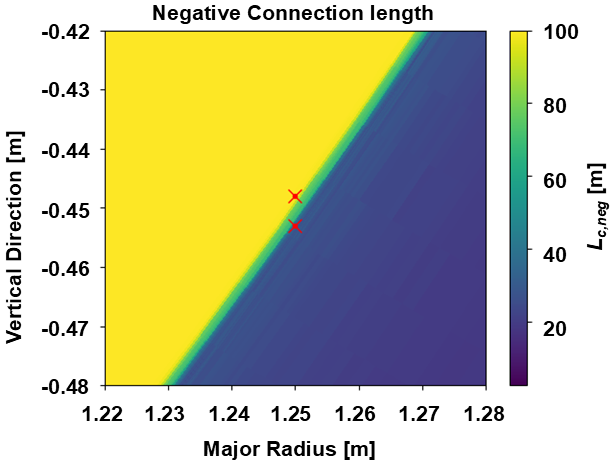}
    \end{subfigure} 
    \caption{Connection length to the divertor following the magnetic field lines in the negative direction up to 100m for the CDN L-mode scenario for the axisymmetric case (left) and RMP cases (right). Red Xs indicating the location for further analysis.}
\label{fig:L-mode_conleng_zoom}
\end{figure}

The first step in this refined analysis is to examine the fluid flow along flux tubes for the L-mode scenario. We analyze two locations in the axisymmetric case and two in the RMP. The two locations in each are shown with red Xs in Figure \ref{fig:L-mode_conleng_zoom}. These plots are connection length plots but with different tracing parameters than before and zoomed in on the plasma edge to highlight a very narrow region of open field lines. The previous connection length analysis was executed with a 1000m tracing limit and the field lines were followed both forward and backward along the magnetic field for a total connection length. For these plots, the field line following limit was set to 100m and only followed in the negative direction. This was found to be the proper tracing limit because, as will be shown later, this is the furthest extent to which open field line behavior occurs along a field line. Plasma further along the line from a target begin to exhibit conditions similar to closed field lines. In these plots, cells with connection length over the limit are again shown in yellow and those with short connection length are in blue. The critical area is then the very thin region with connection length between these shown in green and teal here. Looking at these plots, it can be seen there is a distinct difference between the RMP case and the axisymmetric, where the region with a negative connection length in the RMP case is wider than that of the axisymmetric.

\begin{figure}[htbp]
\centering
    \begin{subfigure}{0.48\textwidth}
        \includegraphics[width=\textwidth]{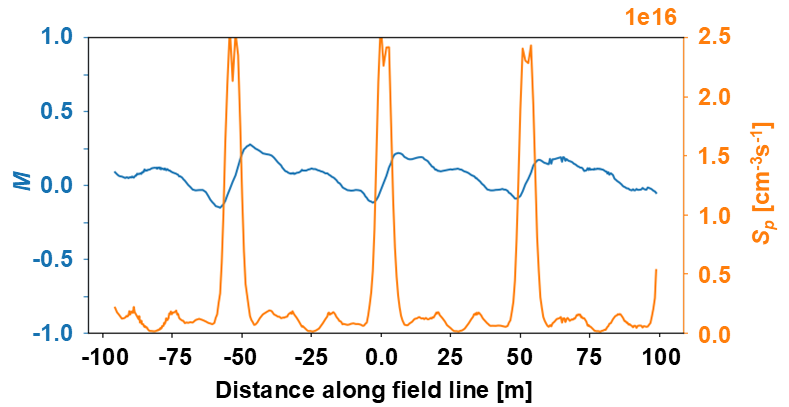}
    \end{subfigure}
    \begin{subfigure}{0.48\textwidth}
        \includegraphics[width=\textwidth]{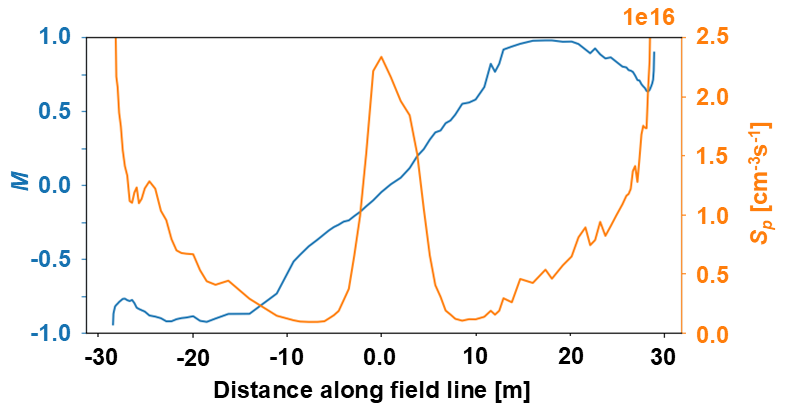}
    \end{subfigure} 
    \caption{Flow speed in terms of Mach number (blue) and ionization rate (orange) plotted along a field line for L-mode axisymmetric case from a point to either of the divertor targets or the limit of the field line tracing. The left plot shows a field line just inside the confined region ($R=1.25$ m, $Z=-0.453$ m) and the right plot shows a field line just outside the confined region ($R=1.25$ m, $Z=-0.455$ m).}
\label{fig:L-mode_noRMP_fieldline}
\end{figure}

Going back to the conditions along the field lines for the points indicated, we will investigate how the plasma behaves on closed versus open field lines. For the axisymmetric case, the locations chosen were $R=1.25$ m, $Z=-0.453$ m, $\phi=0^\circ$ and $R=1.25$ m, $Z=-0.455$ m, $\phi=0^\circ$. These points were picked because they occur in the area of peak ionizations in the edge, and the upper point, $Z=-0.453$ m is located just inside the separatrix, while $Z=-0.455$ m is located just outside the separatrix. The traces for both the ionization rate and flow speed along the field line for these two points are plotted in Figure \ref{fig:L-mode_noRMP_fieldline} where flow speed is plotted in blue in terms of the Mach number $M$ and ionization rate $S_p$ is displayed in orange. For these plots, the $x$-axis is the distance along the field line in the forward direction for positive $x$ and the backward direction for negative $x$, and the starting point is $x=0$.

The left plot in Figure \ref{fig:L-mode_noRMP_fieldline} shows these values along the field line until it hits the computational following limit of 150 m (this is a different field line following limit to better illustrate the behavior along closed field lines). The large peaks in ionization rate in orange indicate when the field line is on the outboard side passing through the particle source and the dips occur where the field line passes many times toroidally around on the inboard side. This field line trajectory is a consequence of the ST configuration and therefore the fueling onto field lines vs. the length scale of the field line to the target is specific to the ST configuration.  There is small, but non-negligible, flow velocity due to the pressure established by the particle source on the outer edge. Plasma particles will tend to flow away from this zone, as seen by the stagnation at the ionization peak, increase in flow away, and then another stagnation point between the sources. Because it is not magnetically connected to the divertor, an ion created on this field line will follow along it either forwards or backwards until cross-field transport causes it to move to another field line.

The point 2mm down from that, across the separatrix, is shown on the right side of the Figure \ref{fig:L-mode_noRMP_fieldline}. This plot indicates how the ionization rate and flow velocity look for the SOL. There is a stagnation point where the peak ionization rate occurs, and then the plasma flows away from the high pressure area and accelerates up as it approaches either the inboard or outboard divertor. There is also a large ionization rate in front of the targets from recycling neutrals, but because of the short connection length and accelerating flow velocity, all of the ions on this field line will tend to flow directly to the target and never fuel the plasma. This has been discussed several times before in this document concerning the argument about the outboard ionization establishing most of the plasma fueling, and is clearly seen here. This finding gives us a physics based way to differentiate if particles ionized in the edge will tend to be able to fuel the plasma or promptly be exhausted to the divertor along the field lines with these near sonic flow levels.

\begin{figure}[htbp]
\centering
    \begin{subfigure}{0.48\textwidth}
        \includegraphics[width=\textwidth]{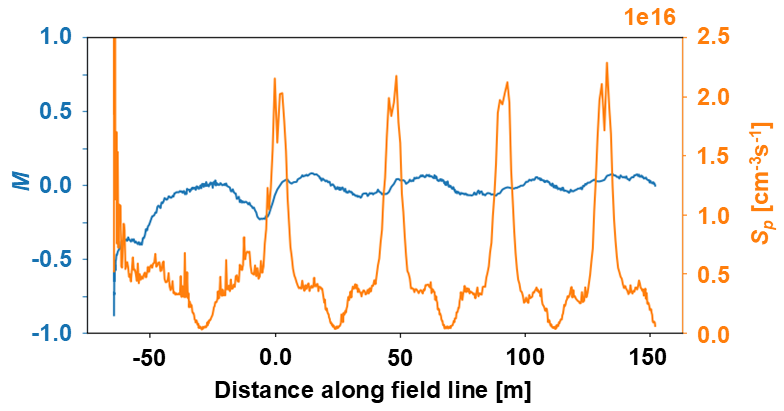}
    \end{subfigure}
    \begin{subfigure}{0.48\textwidth}
        \includegraphics[width=\textwidth]{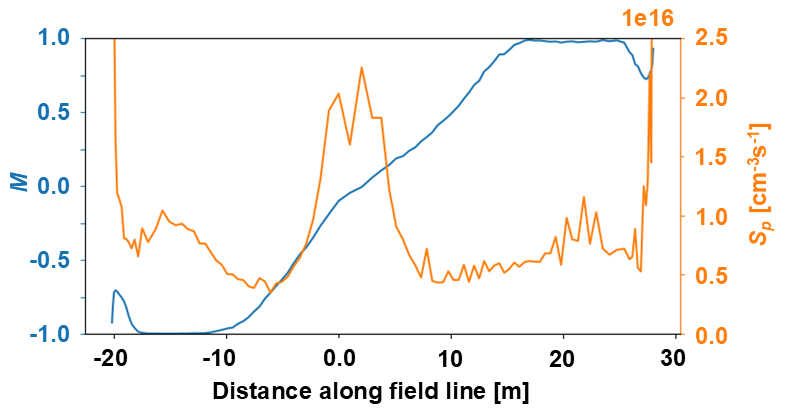}
    \end{subfigure} 
    \caption{Flow speed in terms of Mach number (blue) and ionization rate (orange) plotted along a field line for L-mode even RMP case from an origin point in the edge ending either at a divertor target or the computational following limit. The left plot shows a field line that is the deepest extent of the modified field lines connected directly to the target ($R=1.25$ m, $Z=-0.4489$ m), and the right plot shows a field line just inside the previous separtrix ($R=1.25$ m, $Z=-0.453$ m).}
\label{fig:L-mode_RMP_fieldline}
\end{figure}

To investigate how the RMP-induced changes in the edge structure impact the fueling in relation to this plasma flow constraint, the same analysis was conducted for similar points around the separatrix in the RMP case in Figure \ref{fig:L-mode_RMP_fieldline}. The lowest point for this case, however, is $R=1.25$ m, $Z=-0.453$ m, which had been inside the unperturbed confined region for the axisymmetric case, and is shown in the right plot of Figure \ref{fig:L-mode_RMP_fieldline}. Again the flow speed in terms of $M$ is shown in blue and the ionization rate $S_p$ in orange. This plot looks nearly identical to the point that was in the SOL for the axisymmetric case. This indicates that ionizations occurring here, a point that previously was located in the confined region, will now occur in a flux tube with finite flow directly to the divertor instead of fueling the plasma.

Moving upward in the plasma from that location, the inner edge of the 100m connection length region is at $R=1.25$ m, $Z=-0.4489$ m, shown by the upper red X of the right plot in Figure \ref{fig:L-mode_conleng_zoom}. The trace along the field line starting at that location is shown in the left plot of Figure \ref{fig:L-mode_RMP_fieldline}. The parallel plasma profile from this deeper starting point also shows modified behavior along the field line. In this case, the behavior of the ionization rate and flow velocity in the forward direction is similar to the confined region shown in Figure \ref{fig:L-mode_noRMP_fieldline}. This would indicate that many of the particles originating from just to the right of the starting location would follow a field line that has properties like the confined region and that they also can diffuse perpendicularly off the field line. This would therefore contribute to the core fueling term. To the left of this starting point however, the behavior looks more similar to the SOL. The plasma accelerates as it approaches the divertor in the negative direction along the magnetic field and therefore, ions created there would be exhausted and not contribute to the core fueling term.  Particles ionized on a field line started from this new radial domain in the perturbed plasma can fuel the plasma if they flow in one direction but will be exhausted if they flow in the other. Both occur on the field line with the same total $L_c$, which shows that using the magnetic connection length alone is not sufficient to differentiated between core and edge fueling with the open chaotic layer present.

Using this information, a new integration of all the fueling can be done. Instead of excluding all fueling that occurs on any field line that is open, now fueling will be excluded based on bulk fluid flow along a field line that would take it directly to a divertor target. This is positive flow in the positive magnetic field direction and negative flow in the negative direction, without the field line passing back through the high-ionization region of the outboard side where a stagnation point and flow reversal occurs. Using this as our criteria for fueling in the confined region, the total fueling for the axisymmetric case is $7.98\cdot10^{21}\:\frac{ions}{s}$ and for the RMP case is $7.02\cdot10^{21}\:\frac{ions}{s}$. This is a reduction of 12\% for the even parity case, when we would expect a 15\% reduction in fueling efficiency. This percentage is significantly closer to what the global models used in the previous sections predict than an approach using only the open field lines to determine regions of fueling loss, which produces a much greater reduction in fueling. These results are summarized in Table \ref{tab:Lmode_consource}.

\begin{table}[]
    \centering
    \begin{tabular}{|c|c|c|c|c|}
    \hline
    Scenario & Configuration & Integrated fueling & Percent diff. & Expected from MRPB\\
    \hline
    \hline
    L-mode & Axisymmetric & $7.98\cdot10^{21}\:\frac{ions}{s}$ & -- & -- \\
    \hline
     & RMP & $7.02\cdot10^{21}\:\frac{ions}{s}$ & $-12\%$ & $-15\%$ \\
    \hline
    H-mode & Axisymmetric & $6.98\cdot10^{20}\: \frac{ions}{s}$ & -- & --\\
    \hline
     & RMP & $5.51\cdot10^{20}\:\frac{ions}{s}$ & $-21\%$ & $-10\%$\\
    \hline
    \end{tabular}
    \caption{Integrated fueling rate for L-mode configurations to the confined plasma region and percent difference to the axisymmetric case}
    \label{tab:Lmode_consource}
\end{table}

By conducting the same more detailed field line tracing analysis as the L-mode cases and excluding any ions born on a field line that flows directly to a divertor target, we can now calculate a total fueling to the confined region for both the axisymmetric case and the RMP case for the H-mode scenario. For the axisymmetric case, the total fueling was $6.98\cdot10^{20}\: \frac{ions}{s}$, and for the RMP case, the total fueling was $5.51\cdot10^{20}\:\frac{ions}{s}$. This is about a 21\% decrease from the axisymmetric case to the RMP case. Looking at the MRPB results, we would expect about a 10\% reduction in fueling efficiency. Therefore, our fueling analysis with the comprehensive 3D code EMC3-EIRENE yields about a factor of two higher fueling level. We suspect that this could be due to the fact that EMC3-EIRENE is a steady-state model while the experiment and MRPB both have transient ELM events as well as the full particle source tracing conducted with EIRENE with temperature and density dependent rates compared to the fixed atomic and molecular data used in the MRPB. A more detailed investigation and comparison of rate coefficients used and their temperature dependence in the steep gradient region of the H-mode pedestal could be conducted to resolve this.

\section{Summary and Conclusion}
\label{sec:con}

In this paper we have discussed how the application of RMPs at MAST on both H-mode and L-mode scenarios can result in a particle pump-out and possible mechanisms for this using a variety of analysis tools. The global particle balance using $D_\alpha$ emission to calculate particle fueling and confinement time $\tau_p$ indicated initially that the pump-out was due to reduced particle confinement and therefore a change in transport within the plasma. This, however, may be a naive approximation, as shown by the subsequent sections. By updating the MRPB balance, we were able to accurately represent the discharges and look at how a changing fueling efficiency may also be a cause of this particle pump-out. Indeed, from this model, a change in $\tau_p$ or in fueling efficiency would look identical. Because of this, a more nuanced approach was needed.

The 3-D modeling code EMC3-EIRENE enabled us to look at how the RMPs change the magnetic structure in the edge of the plasma and how that in turn might change the effectiveness of neutral fueling. By following field lines in the edge for the axisymmetric cases, both inside and outside the confined region, one can see how the fueling and ion flow behave on a closed field line and characteristics of the traditional SOL. The application of RMPs resulted in regions of the plasma that were previously on closed field lines that now have direct connection to a divertor target, and field line following in this region showed a change in the ion flow and fueling. The RMPs caused and increased volume with that exhibited conditions similar to the traditional SOL with a short connection to the divertor and flow towards the target. This indicates that fueling in these locations will not effectively fuel the core plasma. With a fueling efficiency change similar to what the MRPB predicts would cause the particle pump-out, this decrease in fueling ionizations is a plausible explanation for the pump-out seen in MAST.

These modeling efforts would be beneficial to apply to current and future experiments at MAST-U, since the new baffle structure allows for novel divertor configurations with significantly different neutral dynamics \cite{moulton_super-x_2024}\cite{harrison_overview_2024}. Using the MRPB model with these new scenarios could be an avenue for estimating the neutral leakage from the baffled divertor area back to the main chamber as well as a way to quantify the fueling efficiency from various puffing locations that the MAST-U can use for main ion fueling \cite{lee_effects_2026}. This could then be extended to run as a predictive model for setting up new scenarios in the device. Additionally, continuing the 3-D field line tracing technique to discharges with RMP applications at MAST-U, as well as other devices, would help confirm this finding that open field lines cause a change in the fueling efficiency. If this effect can be seen in other experiments, that could inform future devices like ITER and pilot plants that will operate with applied RMPs.

\section{Acknowledgements}
This work was funded in part by the U.S. Department of Energy under grants DE-SC0012315 and DE-SC0023381, the Department of Nuclear Engineering and Engineering Physics Goetz fellowship, and the Thomas and Suzanne Werner Chair Professorship and has been carried out within the framework of the EUROfusion Consortium.

\bibliographystyle{iopart-num}

\bibliography{references}

@article{frerichs_volumetric_2021,
	title = {Volumetric recombination in {EMC3}-{EIRENE}: {Implementation} and first application to the pre-fusion power operation phase in {ITER}},
	volume = {28},
	issn = {1070-664X, 1089-7674},
	shorttitle = {Volumetric recombination in {EMC3}-{EIRENE}},
	url = {https://pubs.aip.org/aip/pop/article/1023653},
	doi = {10.1063/5.0062248},
	language = {en},
	number = {10},
	urldate = {2023-06-06},
	journal = {Physics of Plasmas},
	author = {Frerichs, H. and Feng, Y. and Bonnin, X. and Pitts, R. A. and Reiter, D. and Schmitz, O.},
	month = oct,
	year = {2021},
	pages = {102503},
}

@article{liang_active_2007,
	title = {Active control of type-{I} edge localized modes on {JET}},
	volume = {49},
	issn = {0741-3335, 1361-6587},
	url = {https://iopscience.iop.org/article/10.1088/0741-3335/49/12B/S54},
	doi = {10.1088/0741-3335/49/12B/S54},
	language = {en},
	number = {12B},
	urldate = {2022-08-11},
	journal = {Plasma Physics and Controlled Fusion},
	author = {Liang, Y and Koslowski, H R and Thomas, P R and Nardon, E and Jachmich, S and Alper, B and Andrew, P and Andrew, Y and Arnoux, G and Baranov, Y and Bécoulet, M and Beurskens, M and Biewer, T and Bigi, M and Crombe, K and De La Luna, E and de Vries, P and Eich, T and Esser, H G and Fundamenski, W and Gerasimov, S and Giroud, C and Gryaznevich, M P and Harting, D and Hawkes, N and Hotchin, S and Howell, D and Huber, A and Jakubowski, M and Kiptily, V and Kreter, A and Moreira, L and Parail, V and Pinches, S D and Rachlew, E and Schmitz, O and Zimmermann, O and {JET-EFDA Contributors}},
	month = dec,
	year = {2007},
	pages = {B581--B589},
}

@article{evans_suppression_2004,
	title = {Suppression of {Large} {Edge}-{Localized} {Modes} in {High}-{Confinement} {DIII}-{D} {Plasmas} with a {Stochastic} {Magnetic} {Boundary}},
	volume = {92},
	issn = {0031-9007, 1079-7114},
	url = {https://link.aps.org/doi/10.1103/PhysRevLett.92.235003},
	doi = {10.1103/PhysRevLett.92.235003},
	language = {en},
	number = {23},
	urldate = {2022-08-11},
	journal = {Physical Review Letters},
	author = {Evans, T. E. and Moyer, R. A. and Thomas, P. R. and Watkins, J. G. and Osborne, T. H. and Boedo, J. A. and Doyle, E. J. and Fenstermacher, M. E. and Finken, K. H. and Groebner, R. J. and Groth, M. and Harris, J. H. and La Haye, R. J. and Lasnier, C. J. and Masuzaki, S. and Ohyabu, N. and Pretty, D. G. and Rhodes, T. L. and Reimerdes, H. and Rudakov, D. L. and Schaffer, M. J. and Wang, G. and Zeng, L.},
	month = jun,
	year = {2004},
	pages = {235003},
}

@article{tamain_edge_2010,
	title = {Edge turbulence and flows in the presence of resonant magnetic perturbations on {MAST}},
	volume = {52},
	issn = {0741-3335, 1361-6587},
	url = {https://iopscience.iop.org/article/10.1088/0741-3335/52/7/075017},
	doi = {10.1088/0741-3335/52/7/075017},
	language = {en},
	number = {7},
	urldate = {2022-07-13},
	journal = {Plasma Physics and Controlled Fusion},
	author = {Tamain, P and Kirk, A and Nardon, E and Dudson, B and Hnat, B and {the MAST team}},
	month = jul,
	year = {2010},
	pages = {075017},
}

@article{suttrop_first_2011,
	title = {First {Observation} of {Edge} {Localized} {Modes} {Mitigation} with {Resonant} and {Nonresonant} {Magnetic} {Perturbations} in {ASDEX} {Upgrade}},
	volume = {106},
	issn = {0031-9007, 1079-7114},
	url = {https://link.aps.org/doi/10.1103/PhysRevLett.106.225004},
	doi = {10.1103/PhysRevLett.106.225004},
	language = {en},
	number = {22},
	urldate = {2020-07-31},
	journal = {Physical Review Letters},
	author = {Suttrop, W. and Eich, T. and Fuchs, J. C. and Günter, S. and Janzer, A. and Herrmann, A. and Kallenbach, A. and Lang, P. T. and Lunt, T. and Maraschek, M. and McDermott, R. M. and Mlynek, A. and Pütterich, T. and Rott, M. and Vierle, T. and Wolfrum, E. and Yu, Q. and Zammuto, I. and Zohm, H.},
	month = jun,
	year = {2011},
	pages = {225004},
}

@article{kirk_resonant_2010,
	title = {Resonant magnetic perturbation experiments on {MAST} using external and internal coils for {ELM} control},
	volume = {50},
	issn = {0029-5515, 1741-4326},
	url = {https://iopscience.iop.org/article/10.1088/0029-5515/50/3/034008},
	doi = {10.1088/0029-5515/50/3/034008},
	language = {en},
	number = {3},
	urldate = {2020-07-31},
	journal = {Nuclear Fusion},
	author = {Kirk, A. and Nardon, E. and Akers, R. and Bécoulet, M. and De Temmerman, G. and Dudson, B. and Hnat, B. and Liu, Y.Q. and Martin, R. and Tamain, P. and Taylor, D. and {the MAST team}},
	month = mar,
	year = {2010},
	pages = {034008},
}

@article{mckee_increase_2013,
	title = {Increase of turbulence and transport with resonant magnetic perturbations in {ELM}-suppressed plasmas on {DIII}-{D}},
	volume = {53},
	issn = {0029-5515, 1741-4326},
	url = {http://stacks.iop.org/0029-5515/53/i=11/a=113011?key=crossref.100ec6eeaffb8361a136823cce3b7a78},
	doi = {10.1088/0029-5515/53/11/113011},
	language = {en},
	number = {11},
	urldate = {2020-01-30},
	journal = {Nuclear Fusion},
	author = {McKee, G.R. and Yan, Z. and Holland, C. and Buttery, R.J. and Evans, T.E. and Moyer, R.A. and Mordijck, S. and Nazikian, R. and Rhodes, T.L. and Schmitz, O. and Wade, M.R.},
	month = nov,
	year = {2013},
	pages = {113011},
}

@article{jakubowski_uence_2014,
	title = {Inﬂuence of {Magnetic} {Perturbations} on {Particle} {Transport} in magnetic fusion devices},
	volume = {47},
	language = {en},
	number = {24},
	journal = {NIFS-1125},
	author = {Jakubowski, M W and Kirk, A and Suttrop, W and Tanaka, K and Viezzer, E and Wolfrum, E and Hidalgo, C and Kaye, S and McKee, G and Mordijck, S and Schmitz, O and Pedersen, T S},
	month = oct,
	year = {2014},
	pages = {8},
}

@article{feng_3d_1999,
	title = {{3D} fluid modelling of the edge plasma by means of a {Monte} {Carlo} technique},
	volume = {266-269},
	issn = {0022-3115},
	url = {http://www.sciencedirect.com/science/article/pii/S0022311598008447},
	doi = {10.1016/S0022-3115(98)00844-7},
	urldate = {2019-05-09},
	journal = {Journal of Nuclear Materials},
	author = {Feng, Y and Sardei, F and Kisslinger, J},
	month = mar,
	year = {1999},
	pages = {812--818},
}

@article{loarte_transient_2007,
	title = {Transient heat loads in current fusion experiments, extrapolation to {ITER} and consequences for its operation},
	volume = {T128},
	issn = {0031-8949, 1402-4896},
	url = {http://stacks.iop.org/1402-4896/2007/i=T128/a=043?key=crossref.ce940a44f221960f379414bc36ba0498},
	doi = {10.1088/0031-8949/2007/T128/043},
	language = {en},
	journal = {Physica Scripta},
	author = {Loarte, A and Saibene, G and Sartori, R and Riccardo, V and Andrew, P and Paley, J and Fundamenski, W and Eich, T and Herrmann, A and Pautasso, G and Kirk, A and Counsell, G and Federici, G and Strohmayer, G and Whyte, D and Leonard, A and Pitts, R A and Landman, I and Bazylev, B and Pestchanyi, S},
	month = mar,
	year = {2007},
	pages = {222--228},
}

@article{maddison_global_2006,
	title = {Global modelling of tank gas density and effects on plasma density control in {MAST}},
	volume = {48},
	issn = {0741-3335, 1361-6587},
	url = {http://stacks.iop.org/0741-3335/48/i=1/a=007?key=crossref.ab87d2a9207be52ca65db89c2aaf3e02},
	doi = {10.1088/0741-3335/48/1/007},
	language = {en},
	number = {1},
	urldate = {2019-04-19},
	journal = {Plasma Physics and Controlled Fusion},
	author = {Maddison, G P and Turner, A and Fielding, S J and You, S},
	month = jan,
	year = {2006},
	pages = {71--107},
}

@article{wilcox_evidence_2016,
	title = {Evidence of {Toroidally} {Localized} {Turbulence} with {Applied} {3D} {Fields} in the {DIII}-{D} {Tokamak}},
	volume = {117},
	issn = {0031-9007, 1079-7114},
	url = {https://link.aps.org/doi/10.1103/PhysRevLett.117.135001},
	doi = {10.1103/PhysRevLett.117.135001},
	language = {en},
	number = {13},
	urldate = {2019-04-19},
	journal = {Physical Review Letters},
	author = {Wilcox, R. S. and Shafer, M. W. and Ferraro, N. M. and McKee, G. R. and Zeng, L. and Rhodes, T. L. and Canik, J. M. and Paz-Soldan, C. and Nazikian, R. and Unterberg, E. A.},
	month = sep,
	year = {2016},
	pages = {135001},
}

@article{liu_toroidal_2013,
	title = {Toroidal modeling of penetration of the resonant magnetic perturbation field},
	volume = {20},
	issn = {1070-664X, 1089-7674},
	url = {http://aip.scitation.org/doi/10.1063/1.4799535},
	doi = {10.1063/1.4799535},
	language = {en},
	number = {4},
	urldate = {2019-04-19},
	journal = {Physics of Plasmas},
	author = {Liu, Yueqiang and Kirk, A. and Sun, Y.},
	month = apr,
	year = {2013},
	pages = {042503},
}

@article{kirk_understanding_2013,
	title = {Understanding edge-localized mode mitigation by resonant magnetic perturbations on {MAST}},
	volume = {53},
	issn = {0029-5515, 1741-4326},
	url = {http://stacks.iop.org/0029-5515/53/i=4/a=043007?key=crossref.8dc0aecf9fd874f391f10ae9dca52928},
	doi = {10.1088/0029-5515/53/4/043007},
	language = {en},
	number = {4},
	urldate = {2019-04-19},
	journal = {Nuclear Fusion},
	author = {Kirk, A. and Chapman, I.T. and Liu, Yueqiang and Cahyna, P. and Denner, P. and Fishpool, G. and Ham, C.J. and Harrison, J.R. and Liang, Yunfeng and Nardon, E. and Saarelma, S. and Scannell, R. and Thornton, A.J. and {the MAST Team}},
	month = apr,
	year = {2013},
	pages = {043007},
}

@article{martin_design_2013,
    title = {Design evolution and integration of the {ITER} in-vessel components},
    volume = {88},
    issn = {0920-3796},
    url = {https://www.sciencedirect.com/science/article/pii/S0920379613000057},
    doi = {https://doi.org/10.1016/j.fusengdes.2013.01.004},
    number = {9},
    journal = {Fusion Engineering and Design},
    author = {Martin, A. and Calcagno, B. and Chappuis, Ph and Daly, E. and Dellopoulos, G. and Furmanek, A. and Gicquel, S. and Heitzenroeder, P. and Jiming, Chen and Kalish, M. and Kim, D.-H. and Khomiakov, S. and Labusov, A. and Loarte, A. and Loughlin, M. and Merola, M. and Mitteau, R. and Polunovski, E. and Raffray, R. and Sadakov, S. and Ulrickson, M. and Zacchia, F. and Fu, Zhang},
    year = {2013},
    pages = {1955--1959},
}

@article{jeon_suppression_2012,
    title = {Suppression of {Edge} {Localized} {Modes} in {High}-{Confinement} {KSTAR} {Plasmas} by {Nonaxisymmetric} {Magnetic} {Perturbations}},
    volume = {109},
    copyright = {http://link.aps.org/licenses/aps-default-license},
    issn = {0031-9007, 1079-7114},
    url = {https://link.aps.org/doi/10.1103/PhysRevLett.109.035004},
    doi = {10.1103/PhysRevLett.109.035004},
    language = {en},
    number = {3},
    urldate = {2026-02-05},
    journal = {Physical Review Letters},
    author = {Jeon, Y. M. and Park, J.-K. and Yoon, S. W. and Ko, W. H. and Lee, S. G. and Lee, K. D. and Yun, G. S. and Nam, Y. U. and Kim, W. C. and Kwak, Jong-Gu and Lee, K. S. and Kim, H. K. and Yang, H. L.},
    month = jul,
    year = {2012},
    pages = {035004},
}

@article{sun_first_2021,
    title = {First demonstration of full {ELM} suppression in low input torque plasmas to support {ITER} research plan using n = 4 {RMP} in {EAST}},
    volume = {61},
    issn = {0029-5515},
    url = {https://doi.org/10.1088/1741-4326/ac1a1d},
    doi = {10.1088/1741-4326/ac1a1d},
    language = {en},
    number = {10},
    urldate = {2026-02-05},
    journal = {Nuclear Fusion},
    author = {Sun, Y. and Ma, Q. and Jia, M. and Gu, S. and Loarte, A. and Liang, Y. and Liu, Y.Q. and Paz-Soldan, C.A. and Wu, X.M. and Xie, P.C. and Ye, C. and Wang, H.H. and Zhao, J.Q. and Guo, W. and He, K. and Li, Y.Y. and Li, G. and Liu, H. and Qian, J. and Sheng, H. and Shi, T. and Wang, Y.M. and Weisberg, D. and Wan, B. and Zang, Q. and Zeng, L. and Zhang, B. and Zhang, L. and Zhang, T. and Zhou, C. and Contributors, EAST},
    month = oct,
    year = {2021},
    pages = {106037},
}

@article{mordijck_radial_2014,
    title = {The radial electric field as a measure for field penetration of resonant magnetic perturbations},
    volume = {54},
    issn = {0029-5515},
    url = {https://doi.org/10.1088/0029-5515/54/8/082003},
    doi = {10.1088/0029-5515/54/8/082003},
    language = {en},
    number = {8},
    urldate = {2026-02-06},
    journal = {Nuclear Fusion},
    author = {Mordijck, S. and Moyer, R.A. and Ferraro, N.M. and Wade, M.R. and Osborne, T.H.},
    month = jun,
    year = {2014},
    pages = {082003},
}

@article{wilcox_modeling_2017,
    title = {Modeling of {3D} magnetic equilibrium effects on edge turbulence stability during {RMP} {ELM} suppression in tokamaks},
    volume = {57},
    issn = {0029-5515},
    url = {https://doi.org/10.1088/1741-4326/aa7bad},
    doi = {10.1088/1741-4326/aa7bad},
    language = {en},
    number = {11},
    urldate = {2026-02-06},
    journal = {Nuclear Fusion},
    author = {Wilcox, R.S. and Wingen, A. and Cianciosa, M.R. and Ferraro, N.M. and Hirshman, S.P. and Paz-Soldan, C. and Seal, S. K. and Shafer, M.W. and Unterberg, E.A.},
    month = jul,
    year = {2017},
    pages = {116003},
}

@article{kim_transition_2023,
    title = {Transition in particle transport under resonant magnetic perturbations in a tokamak},
    volume = {63},
    issn = {0029-5515},
    url = {https://doi.org/10.1088/1741-4326/acef3c},
    doi = {10.1088/1741-4326/acef3c},
    language = {en},
    number = {10},
    urldate = {2026-02-06},
    journal = {Nuclear Fusion},
    author = {Kim, S.K. and Logan, N.C. and Becoulet, M. and Hoelzl, M. and Hu, Q. and Huijsmans, G.T.A. and Pamela, S.J.P. and Yu, Q. and Yang, S.M. and Paz-soldan, C. and Kolemen, E. and Park, J.-K.},
    month = sep,
    year = {2023},
    pages = {106013},
}

@article{lee_observation_2025,
    title = {Observation of edge kink-like modes induced by resonant magnetic perturbations in {KSTAR} plasmas and their effects on density pump-out},
    volume = {32},
    issn = {1070-664X},
    url = {https://doi.org/10.1063/5.0237640},
    doi = {10.1063/5.0237640},
    number = {1},
    urldate = {2026-02-19},
    journal = {Physics of Plasmas},
    author = {Lee, J. K. and Seol, J. and Lee, H. H. and Liu, Y. Q. and Lee, S. G. and Lee, J. and Kim, B. and Lee, Y. H.},
    month = jan,
    year = {2025},
    pages = {012303},
}

@article{zohm_edge_1996,
    title = {Edge localized modes ({ELMs})},
    volume = {38},
    issn = {0741-3335},
    url = {https://doi.org/10.1088/0741-3335/38/2/001},
    doi = {10.1088/0741-3335/38/2/001},
    language = {en},
    number = {2},
    urldate = {2026-04-14},
    journal = {Plasma Physics and Controlled Fusion},
    author = {Zohm, H.},
    month = feb,
    year = {1996},
    pages = {105},
}

@article{moulton_super-x_2024,
    title = {Super-{X} and conventional divertor configurations in {MAST}-{U} ohmic {L}-mode; a comparison facilitated by interpretative modelling},
    volume = {64},
    issn = {0029-5515},
    url = {https://doi.org/10.1088/1741-4326/ad4f9c},
    doi = {10.1088/1741-4326/ad4f9c},
    language = {en},
    number = {7},
    urldate = {2026-04-14},
    journal = {Nuclear Fusion},
    author = {Moulton, D. and Harrison, J.R. and Xiang, L. and Ryan, P.J. and Kirk, A. and Verhaegh, K. and Wijkamp, T.A. and Federici, F. and Clark, J.G. and Lipschultz, B.},
    month = jun,
    year = {2024},
    pages = {076049},
}

@article{loarte_new_2025,
    title = {The new {ITER} baseline, research plan and open {R}\&{D} issues},
    volume = {67},
    issn = {0741-3335},
    url = {https://doi.org/10.1088/1361-6587/add9c9},
    doi = {10.1088/1361-6587/add9c9},
    language = {en},
    number = {6},
    urldate = {2026-04-16},
    journal = {Plasma Physics and Controlled Fusion},
    author = {Loarte, A and Pitts, R A and Wauters, T and Nunes, I and de Vries, P and Kim, S H and Köchl, F and Polevoi, A and Lehnen, M and Artola, J and Jachmich, S and Pshenov, A and Bai, X and Carvalho, I S and Dubrov, M and Gribov, Y and Schneider, M and Zabeo, L and Bonnin, X and Pinches, S D and Poli, F and Lopez, G Suarez and Merola, M and Escourbiac, F and Hunt, R and Chen, L and Boilson, D and Veltri, P and Casal, N and Preynas, M and Mukherjee, A and Helou, W and Kazarian, F and Willms, S and Bonnet, I and Michling, R and Giancarli, L and van der Laan, J and Walsh, M and Udintsev, V and Reichle, R and Vayakis, G and Fossen, A and Turnyanskiy, M and Becoulet, A and Kamada, Y and Zhuang, G and Xu, G and Gong, X and Huang, J and Jia, M and Ding, R and Qian, J and Sun, Y and Yang, Q and Zhang, L and Xu, M and Zhang, L and Brezinsek, S and Stober, J and Hobirk, J and Rimini, F and Garcia, J and Rao, S L and Ghosh, J and Sharma, D and Magesh, B and Bhattacharya, R P and Matsunaga, G and Urano, H and Hirose, T and Ogawa, K and Motojima, G and Sung, C K and Lee, H H and Park, J K and Cheon, M S and Jeon, Y M and Konovalov, S and Lebedev, S and Kirneva, N and Kashchuk, Y and Bakharev, N and Chen, X and Bortolon, A and Casali, L and Maingi, R and Turco, F and Schmid, K and Liu, Y and Martín-Solís, J R and Angioni, C and Pusztai, I and Fajardo, D and Mateev, D and Lerche, E and van Eester, D and Vincenzi, P and Futtersack, R and Bobkov, V and Colas, L},
    month = jun,
    year = {2025},
    pages = {065023},
}

@article{loarte_effects_2001,
    title = {Effects of divertor geometry on tokamak plasmas},
    volume = {43},
    issn = {0741-3335},
    url = {https://doi.org/10.1088/0741-3335/43/6/201},
    doi = {10.1088/0741-3335/43/6/201},
    language = {en},
    number = {6},
    urldate = {2026-04-16},
    journal = {Plasma Physics and Controlled Fusion},
    author = {Loarte, Alberto},
    month = jun,
    year = {2001},
    pages = {R183},
}

@article{kallenbach_impurity_2013,
    title = {Impurity seeding for tokamak power exhaust: from present devices via {ITER} to {DEMO}},
    volume = {55},
    issn = {0741-3335},
    shorttitle = {Impurity seeding for tokamak power exhaust},
    url = {https://doi.org/10.1088/0741-3335/55/12/124041},
    doi = {10.1088/0741-3335/55/12/124041},
    language = {en},
    number = {12},
    urldate = {2026-04-16},
    journal = {Plasma Physics and Controlled Fusion},
    author = {Kallenbach, A and Bernert, M and Dux, R and Casali, L and Eich, T and Giannone, L and Herrmann, A and McDermott, R and Mlynek, A and Müller, H W and Reimold, F and Schweinzer, J and Sertoli, M and Tardini, G and Treutterer, W and Viezzer, E and Wenninger, R and Wischmeier, M and Team, the ASDEX Upgrade},
    month = nov,
    year = {2013},
    pages = {124041},
}

@article{casali_improved_2020,
    title = {Improved core-edge compatibility using impurity seeding in the small angle slot ({SAS}) divertor at {DIII}-{D}},
    volume = {27},
    issn = {1070-664X},
    url = {https://doi.org/10.1063/1.5144693},
    doi = {10.1063/1.5144693},
    number = {6},
    urldate = {2026-04-16},
    journal = {Physics of Plasmas},
    author = {Casali, L. and Osborne, T. H. and Grierson, B. A. and McLean, A. G. and Meier, E. T. and Ren, J. and Shafer, M. W. and Wang, H. and Watkins, J. G.},
    month = jun,
    year = {2020},
    pages = {062506},
}

@article{turco_physics_2023,
    title = {The physics basis to integrate an {MHD} stable, high-power hybrid scenario to a cool divertor for steady-state reactor operation},
    volume = {63},
    issn = {0029-5515},
    url = {https://doi.org/10.1088/1741-4326/acb370},
    doi = {10.1088/1741-4326/acb370},
    language = {en},
    number = {3},
    urldate = {2026-04-16},
    journal = {Nuclear Fusion},
    author = {Turco, F. and Petrie, T. and Osborne, T. and Petty, C.C. and Luce, T.C. and Grierson, B. and Odstrcil, T. and Van Zeeland, M.A. and Liu, D. and Casali, L. and Boyes, W. and Smith, S.P. and Shen, H. and Kostuk, M. and Brennan, D.},
    month = feb,
    year = {2023},
    pages = {036020},
}

@article{kallenbach_partial_2015,
    title = {Partial detachment of high power discharges in {ASDEX} {Upgrade}},
    volume = {55},
    issn = {0029-5515},
    url = {https://doi.org/10.1088/0029-5515/55/5/053026},
    doi = {10.1088/0029-5515/55/5/053026},
    language = {en},
    number = {5},
    urldate = {2026-04-16},
    journal = {Nuclear Fusion},
    author = {Kallenbach, A. and Bernert, M. and Beurskens, M. and Casali, L. and Dunne, M. and Eich, T. and Giannone, L. and Herrmann, A. and Maraschek, M. and Potzel, S. and Reimold, F. and Rohde, V. and Schweinzer, J. and Viezzer, E. and Wischmeier, M. and Team, the ASDEX Upgrade},
    month = apr,
    year = {2015},
    pages = {053026},
}

@article{wagner_quarter-century_2007,
    title = {A quarter-century of {H}-mode studies},
    volume = {49},
    issn = {0741-3335},
    url = {https://doi.org/10.1088/0741-3335/49/12B/S01},
    doi = {10.1088/0741-3335/49/12B/S01},
    language = {en},
    number = {12B},
    urldate = {2026-04-16},
    journal = {Plasma Physics and Controlled Fusion},
    author = {Wagner, F},
    month = nov,
    year = {2007},
    pages = {B1},
}

@article{casali_achievement_2025,
    title = {Achievement of highly radiating plasma in negative triangularity and effect of reactor-relevant seeded impurities on confinement and transport},
    volume = {67},
    issn = {0741-3335},
    url = {https://doi.org/10.1088/1361-6587/ada1ca},
    doi = {10.1088/1361-6587/ada1ca},
    language = {en},
    number = {2},
    urldate = {2026-04-16},
    journal = {Plasma Physics and Controlled Fusion},
    author = {Casali, L and Eldon, D and Odstrcil, T and Mattes, R and Welsh, A and Lee, K and Nelson, A O and Paz-Soldan, C and Khabanov, F and Cote, T and McLean, A G and Scotti, F and Thome, K E},
    month = jan,
    year = {2025},
    pages = {025007},
}

@article{casali_neutral_2020,
    title = {Neutral leakage, power dissipation and pedestal fueling in open vs closed divertors},
    volume = {60},
    issn = {0029-5515},
    url = {https://doi.org/10.1088/1741-4326/ab8d06},
    doi = {10.1088/1741-4326/ab8d06},
    language = {en},
    number = {7},
    urldate = {2026-04-18},
    journal = {Nuclear Fusion},
    author = {Casali, L. and Eldon, D. and Boedo, J.A. and Leonard, T. and Covele, B.},
    month = jun,
    year = {2020},
    pages = {076011},
}

@article{harrison_overview_2024,
    title = {Overview of physics results from {MAST} upgrade towards core-pedestal-exhaust integration},
    volume = {64},
    issn = {0029-5515},
    url = {https://dx.doi.org/10.1088/1741-4326/ad6011},
    doi = {10.1088/1741-4326/ad6011},
    language = {en},
    number = {11},
    urldate = {2025-07-23},
    journal = {Nuclear Fusion},
    author = {Harrison, J.R. and Aboutaleb, A. and Ahmed, S. and Aljunid, M. and Allan, S.Y. and Anand, H. and Andrew, Y. and Appel, L.C. and Ash, A. and Ashton, J. and Bachmann, O. and Barnes, M. and Barrett, B. and Baver, D. and Beckett, D. and Bennett, J. and Berkery, J. and Bernert, M. and Boeglin, W. and Bowman, C. and Bradley, J. and Brida, D. and Browning, P.K. and Brunetti, D. and Bryant, P. and Bryant, J. and Buchanan, J. and Bulmer, N. and Carruthers, A. and Casali, L. and Cecconello, M. and Chen, Z.P. and Clark, J. and Cowley, C. and Coy, M. and Crocker, N. and Cunningham, G. and Cziegler, I. and Da Assuncao, T. and Damizia, Y. and Davies, P. and Day, I.E. and Derks, G.L. and Dixon, S. and Doyle, R. and Dreval, M. and Dunne, M. and Duval, B.P. and Eagles, T. and Edmond, J. and El-Haroun, H. and Elmore, S.D. and Enters, Y. and Faitsch, M. and Federici, F. and Fedorczak, N. and Felici, F. and Field, A.R. and Fitzgerald, M. and Fitzgerald, I. and Fitzpatrick, R. and Frassinetti, L. and Fuller, W. and Gahle, D. and Galdon-Quiroga, J. and Garzotti, L. and Gee, S. and Gheorghiu, T. and Gibson, S. and Gibson, K.J. and Giroud, C. and Greenhouse, D. and Hall-Chen, V.H. and Ham, C.J. and Harrison, R. and Henderson, S.S. and Hickling, C. and Hnat, B. and Howlett, L. and Hughes, J. and Hussain, R. and Imada, K. and Jacquet, P. and Jepson, P. and Kandan, B. and Katramados, I. and Kazakov, Y.O. and King, D. and King, R. and Kirk, A. and Knolker, M. and Kochan, M. and Kogan, L. and Kool, B. and Kotschenreuther, M. and Lee, K. -W. and Lees, M. and Leonard, A.W. and Liddiard, G. and Lipschultz, B. and Liu, Y.Q. and Lomanowski, B.A. and Lonigro, N. and Lore, J. and Lovell, J. and Mahajan, S. and Maiden, F. and Man-Friel, C. and Mansfield, F. and Marsden, S. and Martin, R. and Mazzi, S. and McAdams, R. and McArdle, G. and McClements, K.G. and McClenaghan, J. and McConville, D. and McKay, K. and McKnight, C. and McKnight, P. and McLean, A. and McMillan, B.F. and McShee, A. and Measures, J. and Mehay, N. and Michael, C.A. and Militello, F. and Morbey, D. and Mordijck, S. and Moulton, D. and Myatra, O. and Nelson, A.O. and Nicassio, M. and O’Mullane, M.G. and Oliver, H.J.C. and Ollus, P. and Osborne, T. and Osborne, N. and Parr, E. and Parry, B. and Patel, B.S. and Payne, D. and Paz-Soldan, C. and Phelps, A. and Piron, L. and Piron, C. and Prechel, G. and Price, M. and Pritchard, B. and Proudfoot, R. and Reimerdes, H. and Rhodes, T. and Richardson, P. and Riquezes, J. and Rivero-Rodriguez, J.F. and Roach, C.M. and Robson, M. and Ronald, K. and Rose, E. and Ryan, P. and Ryan, D. and Saarelma, S. and Sabbagh, S. and Sarwar, R. and Saunders, P. and Sauter, O. and Scannell, R. and Schuett, T. and Seath, R. and Sharma, R. and Shi, P. and Sieglin, B. and Simmonds, M. and Smith, J. and Smith, A. and Soukhanovskii, V. A. and Speirs, D. and Staebler, G. and Stephen, R. and Stevenson, P. and Stobbs, J. and Stott, M. and Stroud, C. and Tame, C. and Theiler, C. and Thomas-Davies, N. and Thornton, A.J. and Tobin, M. and Vallar, M. and Vann, R.G.L. and Velarde, L. and Verhaegh, K. and Viezzer, E. and Vincent, C. and Voss, G. and Warr, M. and Wehner, W. and Wiesen, S. and Wijkamp, T.A. and Wilkins, D. and Williams, T. and Wilson, T. and Wilson, H.R. and Wong, H. and Wood, M. and Zamkovska, V.},
    month = aug,
    year = {2024},
    note = {Publisher: IOP Publishing},
    pages = {112017},
}

@article{lee_effects_2026,
    title = {The effects of gas puff locations and divertor closure on detachment conditions in {MAST}-{U}},
    volume = {66},
    issn = {0029-5515},
    url = {https://doi.org/10.1088/1741-4326/ae3625},
    doi = {10.1088/1741-4326/ae3625},
    language = {en},
    number = {2},
    urldate = {2026-04-18},
    journal = {Nuclear Fusion},
    author = {Lee, K. and Casali, L. and Smiskey, J. and Moulton, D. and Ryan, P. and Lonigro, N.},
    month = jan,
    year = {2026},
    pages = {026047},
}

@article{harrison_characteristics_2014,
    title = {Characteristics of {X}-point lobe structures in single-null discharges on {MAST}},
    volume = {54},
    issn = {0029-5515},
    url = {https://doi.org/10.1088/0029-5515/54/6/064015},
    doi = {10.1088/0029-5515/54/6/064015},
    language = {en},
    number = {6},
    urldate = {2026-06-02},
    journal = {Nuclear Fusion},
    author = {Harrison, J.R. and Kirk, A. and Chapman, I.T. and Cahyna, P. and Liu, Yueqiang and Nardon, E. and Thornton, A.J.},
    month = may,
    year = {2014},
    pages = {064015},
}

\end{document}